\documentclass[preprint,showkeys,preprintnumbers,amsmath,amssymb,showpacs]{revtex4}

\usepackage{verbatim}
\usepackage{graphicx}
\usepackage{dcolumn}
\usepackage{bm}

\newcommand{\expect}[1]{\langle #1 \rangle}

\begin{document}

\title{Single-layer metal-on-metal islands driven by strong time-dependent forces} 

\author{Janne Kauttonen}
 \affiliation{Department of Physics, University of Jyv\"askyl\"a, P.O. Box 35, FI-40014 Jyv\"askyl\"a, Finland}
 \email{janne.kauttonen@jyu.fi}
\author{Juha Merikoski}
 \affiliation{Department of Physics, University of Jyv\"askyl\"a, P.O. Box 35, FI-40014 Jyv\"askyl\"a, Finland}  
  
\date{\today}

\begin{abstract}

Non-linear transport properties of single-layer metal-on-metal islands driven with strong static and time-dependent forces are studied.  We apply a semi-empirical lattice model and use master equation and kinetic Monte Carlo simulation methods to compute observables such as the velocity and the diffusion coefficient.  Two types of time-dependent driving are considered: a pulsed rotated field and an alternating field with a zero net force (electrophoretic ratchet). Small islands up to 12 atoms were studied in detail with the master equation method and larger ones with simulations.  Results are presented mainly for a parametrization of Cu on Cu(001) surface, which has been the main system of interest in several previous studies.  The main results are that the pulsed field can increase the current in both diagonal and axis direction when compared to static field, and there exists a current inversion in the electrophoretic ratchet.  Both of these phenomena are a consequence of the coupling of the internal dynamics of the island with its transport.  In addition to the previously discovered ``magic size'' effect for islands in equilibrium, a strong odd-even effect was found for islands driven far out of equilibrium.  Master equation computations revealed non-monotonous behavior for the leading relaxation constant and effective Arrhenius parameters.  Using cycle optimization methods, typical island transport mechanisms are identified for small islands.

\end{abstract}


\maketitle

\section{Introduction}

Research of transport of complex molecular and micro-scale objects has flourished in the last two decades.  Important discoveries and knowledge has been gained especially on molecular motors and in surface physics, where the development in experimental and computational techniques have reached the level to allow studying and manipulation of individual molecules and atoms \cite{Marchesoni,nissila_review}.  Diffusion of adsorbates on surfaces is perhaps the most elementary transport process occurring on surfaces \cite{pintakirja}.  It is crucial for more complex surface phenomena such as crystal growth, associative desorption, heterogeneous catalysis and chemical reactions.  In this work, we study the properties of single-layer atom islands under the effect of strong and time-dependent external forces.

During 90s and early 2000s, lots of insight of the microscopic details for single-layer metal-on-metal islands were revealed by not only experiments, but also Monte Carlo and molecular dynamic simulations.  Today equilibrium properties of single-layer islands and voids are well known, such as energetics of single and many-atom processes, scaling of the diffusion coefficient and the ``magic size'' effect for small islands \cite{nissila_review,Yufen,Voter,Merikoski2}.  Near-equilibrium properties have been studied for small electromigration forces in the linear response regime and phenomena such as oscillatory motion and deformations of islands have been found \cite{Karimi1,Bitar,Rusanen,Rous}.  Also some properties under strong forces have been studied with continuum models \cite{jatkumomallit}.  Using a discrete atomistic model on a square lattice, we carry out a detailed study of the non-linear properties arising with strong electromigration-type forces.

The motivation for this work is two-fold. First, the properties of complex nonequilibrium systems are theoretically interesting.  Many-particle systems often exhibit surprising and unexpected behavior when driven far from equilibrium.  A good example of this is the \emph{ratchet effect} \cite{Reimann}, where time-dependent external forces and many-particle interactions can cause surprising non-linear effects such as current inversions \cite{monta_partikkelia,Kauttonen1,Kulakowski}.  Due to the lack of general theory of nonequilibrium statistical physics, the principal way to gain insight of these systems is through numerical methods.  The island diffusion model offers strongly correlated many-particle interactions with transitions occurring at several time-scales, which makes it a good prototype for further study.  Second, although experimentally realizable electromigration forces in most cases are too small to cause significant non-linear effects alone, combined with other methods that can decrease energy barriers, such as strain, manipulation with scanning tunneling microscopy and electric fields \cite{strain,STM}, non-linear effects are expected to emerge.  In this article we give a comprehensive look at the origins of these effects.

Our main interest lies in the generic steady state transport properties of two-dimensional islands on a solid surface.  To keep the model as simple and general as possible, we apply a well studied model of hopping atoms in the square lattice with nearest-neighbor energetics \cite{Merikoski1}.  Drift is generated by adding an external time-dependent electromigration-type force that causes biased hopping of atoms.  We are particularly interested in the behavior of small islands with up to 20 atoms where the discreteness and finite-size effects are strongly present.  For these smaller islands, entropic effects are not yet dominant.

In comparison to previous works, where islands driven by static fields have been considered, we also consider time-dependent fields where the direction and amplitude of the field is temporally varied.  For such fields island transport becomes also frequency dependent.  Due to the interplay between island configurations and strong time-dependent forces, one could expect phenomena such as current inversion or increase to appear (see e.g.~Ref.~\cite{Kulakowski,Xie}).
The pulsed field and the electrophoretic ratchet have been previously studied especially within the electrophoresis community to enhance the efficiency of separating DNA molecules \cite{electrophoresis}.

We present numerical results for a model of Cu islands on the Cu(001) surface, which is one of the most studied systems in the literature on monolayer islands.  Because of the square lattice geometry, the (001) surface is a good candidate for investigating the effects of rotated field, which would be less pronounced for e.g.~the (111) surfaces.  We apply both the master equation (ME) method and kinetic Monte Carlo (MC) simulations.  Both of these methods have been separately used in several previous studies \cite{hiukkasmallit,Karimi1,Evans,Voter,Skodje,Rusanen,Mattsson,Rous,haihtuminen,Merikoski2,Trushin}.  We apply them both in order to utilize their strengths and also compare their differences.  Being numerically exact, the ME method allows an accurate evaluation of some elusive properties, such as the leading relaxation constant and the Arrhenius parameters, that cannot be easily obtained from simulations.  On the other hand, the MC method can handle larger and more complicated systems with the expense of being numerically less accurate.  Although the parameter space for the model is extensive, we find several generic features of transport.

This Article is organized as follows.  In Section II the notation, model and applied computational methods are described, and in Sections III-V computational results are shown for time-independent and time-dependent fields, respectively.  In Section VI we present our conclusions and discuss the generality and applicability of our main conclusions.

\section{Model and numerical methods}

The model consists of two-dimensional atom islands on an unbounded square-lattice surface.  In this set-up, each atom has up to four nearest and four next-nearest (diagonal) neighbors.  To keep the islands unbroken, we require that each atom must be connected to the island with at least one diagonal neighbor.  The dynamics is created by single atom hops in continuous time with rates given by a semi-empirical model parameterized by using embedded atom method \cite{Merikoski1}.  Within this model, the changes in the binding energy for an atomic transition is computed from the change in nearest neighbor atom count.  Despite its simplicity, the energetics given by the model are in good agreement with molecular dynamical computations (see \cite{Sukky} for the most recent results).  The transition rate $\Gamma_{i,f}$ from the initial state (i) to the final state (f) is given by
\begin{equation}
\Gamma_{f,i}(t) = \nu \exp \left( \frac{-E_{S} - \max \left\{ 0,E_B \Delta_{f,i} \right\} + E_{f,i}(t) )}{k_{\rm B} T} \right),
\label{eq:taajuudet}
\end{equation}
where $\nu$ is an effective vibrational frequency, $E_{\rm S}$ is the energy barrier for the atom transition along the island edge, $E_{{\rm B}} \Delta_{f,i}$ ($\Delta_{f,i} = -3, \dots ,3$) is the change in the binding energy with nearest neighbor bonds, and $E_{f,i}(t)$ gives the time-dependent contribution (positive or negative) to the transition rate by an external electromigration-type force and depends on the magnitudes and directions of the field and atom displacement.  For Cu on Cu(001) we use $E_{\rm B}=0.260~{\rm eV}$ and $E_{\rm S} = 0.258 ~{\rm eV}$ \cite{Merikoski1}.  We set $\nu$ and the lattice constant to 1 for the rest of the paper.  Since the barrier $E_{\rm S}$ appears in all transitions, it can be integrated in the prefactor by defining a new temperature-dependent prefactor $\widetilde{\nu}:=\nu \exp (-E_{S}/k_{\rm B} T )$.  In the zero-field case (i.e.~$E_{f,i} \equiv 0$), the probability of a configuration in the steady state is proportional to $\exp(-L_i E_{\rm B} / 2 k_{\rm B} T)$, where $L_i$ is the length of the island perimeter in configuration $i$. Due to the large separation of the energy barriers, there are at temperatures $T<1000~{\rm K}$ four well separated microscopic rate 
parameters in the system in zero field and up to six for large fields ($E \sim E_B$).

Although this simplified model is not microscopically accurate, it captures the key elements of the dynamics, respects the detailed balance condition (for $E_{f,i}(t) \equiv 0$) to avoid spurious currents and is straightforward to apply in computations.  We want to apply a simplified model as we do not wish to study only a particular system but rather to investigate properties which should not depend on the details of atom-atom interaction. Therefore, we use a simple kinetic model containing as few parameters as possible.  Experimentally realizable field amplitudes in electromigration are of order $E \sim 10^{-5}~{\rm eV}$ \cite{Karimi1,Rous}.  Because of this, transport properties of non-continuous islands have been previously studied only within the linear-response regime with very small fields ($E<0.01~{\rm eV}$).

As the detachment of atoms from the island is denied in our model, the number of atoms in the island remains a constant.  This is a well-justified approximation for small fields (i.e.~$E_{f,i} \ll E_{B}$) and temperatures far below the melting point of the metal ($1358~{\rm K}$ for Cu).  Also, if the density of free surface atoms is such that evaporation and condensation are balanced, the island size could be kept constant on average also with unrestricted dynamics. When the detachment/attachment processes are rare compared to other processes, the properties of the variable size islands follow from those of the fixed size islands because the island size remains fixed for long periods of time, thus allowing relaxation between events.  In such cases, for example the mean velocity of the variable size island under a driving force is a combination of velocities of fixed size islands over a wide time scale.  Only when the detachment/attachment processes become very frequent, which necessarily occurs at high temperatures and very strong fields, this commonly used picture of noninteracting islands is no longer valid.  On the basis of previous studies at (or near) equilibrium (e.g.~\cite{haihtuminen,Merikoski2}) and the results of Sections III-IV, at least within temperatures below $800~{\rm K}$ and $E<0.1~{\rm eV}$ one can assume that the dynamics is not dominated by the detachment/attachment processes and the transport is mainly caused by the periphery diffusion.

As we are interested in transport properties, the first interesting observable is the velocity $\vec{v}$ for the center of mass of the island, defined as
\begin{equation*}
\vec{v} = \lim_{t \rightarrow \infty} \frac{1}{t} \expect{\vec{x}(t)} 
\end{equation*}
where $\vec{v}=(v_x,v_y)$ and $\vec{x} = \left( x, y \right)$ is the center of mass position at the surface using the main axes of the lattice.  We define the measuring direction by angle $\gamma$, i.e.~we measure $v = v_x \cos(\gamma) + v_y \sin(\gamma)$.  The field is defined by the amplitude $E\geq0$ and the angle $\alpha$ of the field direction, i.e.~$\vec{E} = \left( E  \cos(\alpha), E \sin(\alpha) \right)$.  

\begin{figure}
\includegraphics[width=5cm]{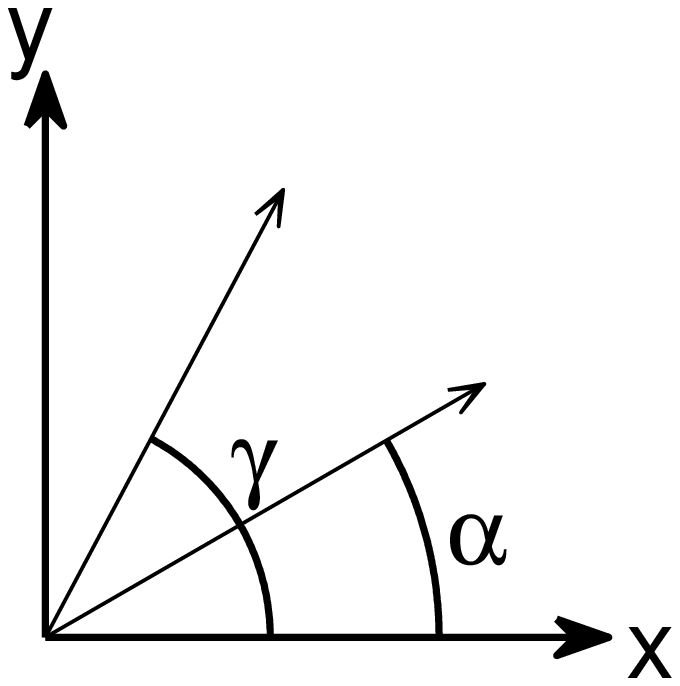}
\includegraphics[width=14cm]{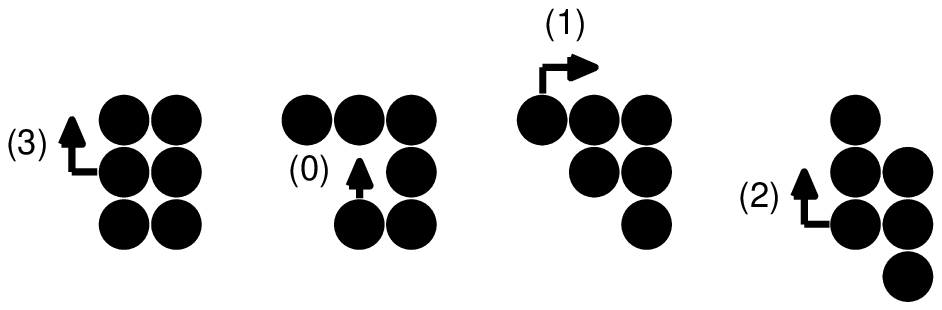}
\caption{Used notation for the angles $\gamma$ and $\alpha$ (see text) and an illustration of the model with a 6-atom ($N=6$) island going through 4 consecutive transitions.  Small arrows indicate the transitions and their corresponding values of the $\Delta_{f,i}$ parameter in Eq.~(1). }
\label{fig:esimerkkikuva1}
\end{figure}

The effective diffusion coefficient $D_{\rm eff}$ in direction $\gamma$ is defined by
\begin{equation*}
D_{\rm eff} = \frac{1}{2} \lim_{t \rightarrow \infty} \frac{1}{t} \left( \expect{x_{\gamma}(t)^2} -  \expect{x_{\gamma}(t)}^2 \right),
\end{equation*}
where $x_{\gamma} = \cos(\gamma) x + \sin (\gamma) y$.  We are also interested in observables determining the geometry of the island, such as the average thickness and length in the direction of an axis.  The observables are computed by kinetic Monte Carlo simulations and solving master equations directly.  The model and its parameters are illustrated in Fig.~\ref{fig:esimerkkikuva1}.

To illustrate atomic transitions and the effect of the field, consider a non-zero field with $\alpha = 90^\circ$ such that $\vec{E}=(0,E)$) in Fig.~\ref{fig:esimerkkikuva1}.  The energy barriers for the four transitions shown are decreased by $E$ and similarly increased for the corresponding inverse transitions.  For $\alpha = 45^\circ$, the barrier of transition (1) is decreased by $2E/\sqrt{2}$ and that of transition (0) by $E/\sqrt{2}$. 
In what we shall call the Monte Carlo (MC) model, two independent separate jumps are required to go around a corner [processes (1)-(3)], whereas in the Master Equation (ME) model such transitions occur by direct diagonal jumps. We next describe these models and the numerical methods best suited for them.

\subsection{Monte Carlo model and method}

We applied the continuous-time Monte Carlo method in the N-fold way \cite{Blue}.  In the presence of the field, there are up to $4 \times 4=16$ different reaction classes in the simulation (4 different nearest-neighbor energy barriers and 4 different hopping directions). In the model used in the simulation, only jumps between nearest-neighbor sites were allowed.

The break-up of the island was prevented by checking the global connectivity of the island each time a potential island-breaking move was tried. This way the dynamics of our MC model is less restricted than in previous studies, where all potential breaking-up transitions were omitted \cite{Merikoski2,Rusanen}.  Global checking however causes some slow-down of the simulations, but in most cases checks need to be performed only once per $10^3 \dots 10^5$ attempted moves.  Only at very high temperatures ($T>1000~{\rm K}$) and/or very strong fields, the tryout frequency of illegal transitions eventually reaches tens of percents.

To gain accurate results for the observables, simulations were performed at temperatures above $400~{\rm K}$ where the island mobility is large enough.  All results were averaged from 100-2000 independent runs (more iterations for small islands).   The velocity and the diffusion coefficient were computed from the slopes of the center of mass of the position, other (non-increasing) observables were averaged within the steady state.  The approach to the steady state was confirmed from the position and geometry data.  Using geometry data, such as the perimeter length and the island width/length, was found to be important, since the actual relaxation observed through the island shape can take significantly longer than it appears from the position data alone.  Initial states for the simulations were sampled from the corresponding equilibrium shapes.  Since these states are generally far from the nonequilibrium steady states, the simulations quickly become difficult for large islands because of the long times needed to reach the steady state.  Also because of the greater migration velocity, simulations with the field direction along the axis tend to be more accurate when compared to the diagonal fields.

\subsection{Master equation model and method}

After the rates are given, the dynamics of the island is governed by the master equations (see e.g.~\cite{Kampen})
\begin{equation}
\frac{d P_y(t)}{dt}= \sum_{y' \neq y} \left[ H_{y,y'}(t) P_{y'}(t) - H_{y',y}(t) P_y(t) \right],
\label{eq:generaattori}
\end{equation}
where $P_y(t)$ is the probability of island configuration $y$ at time $t$ and $H_{y,y'}(t) := \Gamma_{y,y'}(t)$ is the stochastic generator of the process.  By setting $H_{y,y} = - \sum_{y' \neq y} H_{y',y}$, above can be written in the matrix form $\dot{P}(t)=H P(t)$.  Let the number of equations be $Y$ (i.e.~dimension of square matrix $H$), which is the number of allowed configurations for the system.  The matrix $H$ can be explicitly build and the dynamics solved for small islands.

To be able to study islands up to $N=12$, we reduced the number of island configurations by allowing only such states that do not include configurations with only diagonally connected atoms or parts of the island. To facilitate going around the corner, which is necessary for long-range transport of the island, we allow direct diagonal jumps like the jumps (1-3) shown in Fig.~\ref{fig:esimerkkikuva1}. In the MC model going around the corner is possible by two jumps.   The energy barrier of the diagonal jump is approximated by a sum of the binding energy difference between the initial and intermediate state and the total energy difference caused by the field.  Further reduction was made by disallowing vacancies inside the islands, which however has only a minor effect on the number of available states.  
This way we have defined the ME model.

These approximations cause only a minor differences between MC and ME models in equilibrium ($E \equiv 0$), where the weights only depend on the total energy of the configuration and the island prefers compact rectangular shapes.  For nonequilibrium states, major differences between the models are expected especially for field amplitudes approaching the binding energy $E \sim E_B$.  This is mainly because of the trap configurations (see Sec.~II.C) and the diagonal jump approximation.  The difference between the MC and ME models depends on how important the corner states are for the dynamics.  

The remaining island configurations are known as polyominos (or lattice animals), and their counting and statistical properties are known at least up to $N<47$ \cite{Jensen}.  Since the practical limit for numerical master equation computations is about $10^6$ states, the largest system studied in this work is the 12-atom island with 468837 states (505861 if vacancies are allowed).  From now on, we call this reduced model the \emph{ME model}, whereas the full model is called the \emph{MC model}.  For a comparison, only up to 8-atom islands could be treated by applying the ME method directly to the full MC model without any above reductions.

The ME method suffers from a low temperature problem as the MC method does.  As the temperature decreases, the difference between the largest and the smallest rate increases, which eventually leads to a very stiff set of linear equations (i.e.~$H$ is badly scaled). This limits the minimum temperature for practical computations to be around $500~{\rm K}$.  When computing the steady state (i.e.~solve $H \vec{x} = 0$), the stability of computations can be somewhat improved by using diagonal pre-conditioning matrix with entries $1/H_{i,i}$ (this works because $H$ is diagonally dominant).  This decreases the lowest reachable temperature to be near $300~{\rm K}$ for the velocity.  However, this procedure was not found to improve computations of the effective diffusion coefficient that requires solving linear equations of the type $H \vec{x} = \vec{b}$.

In general, reaching low temperatures with stable numerics would require coarse graining techniques to reduce the separation of rates, and it is a possible continuation for this work \footnote{This kind of idea has been applied in \cite{Evans}, where low temperature coarse graining is carried on by hand for islands $N=3 \dots 5$.}.

\subsubsection{Building the stochastic generators}

Master equation computations are carried out as follows. First all the allowed island configurations are enumerated with a brute force method \cite{Redelmeier}.  After enumeration, elements of the stochastic matrix $H$ are found by finding all allowed transitions between the configurations.  This is the most time-consuming part of the computations as in the general case it scales as $Y^2$ (comparing all configurations against each other).  However, for the current model we can take advantage of that only single atom transitions are allowed (for general transitions some pattern recognition algorithm would be required).  Therefore we sort the states by their projection along coordinate axes and their diagonals, which greatly reduces the number of states that need to be compared against each other.  We ended up with matrices that include rate classes and displacements for transitions.  These matrices are stored.  Before numerically solving the master equations, the final matrix is created by putting the actual rates into to the matrix. This final step is fast since the matrices are very sparse.  Most parts of these computations can be easily parallelized by dividing the matrices in smaller parts.  To obtain the velocity and the diffusion coefficient, we apply numerical linear algebra methods described in  Ref.~\cite{Kauttonen1}.

Finally we note that there is indeed a much more efficient way to enumerate polyominos by using transfer matrix method \cite{Jensen}, but since we need to really build the stochastic generators instead of just enumerating, this approach is not useful here.  Also since only small systems can be treated (here $N<13$), the time consumption of the enumerating process is negligible compared to other parts of the computation.

\subsubsection{Transition paths of the islands}

Transport of the center of mass of the island is caused by sequences of single atom hops i.e.~transition paths.  One can try to find paths via different methods such as DFT computations, simulations and also by pure reasoning for small systems (see e.g.~\cite{Evans,Voter}).  However, since the ME method takes into account all possible states and transitions, one can utilize a method from the graph analysis known as \emph{mean cycle optimization} \cite{Dasdan}. The optimal mean cycle is a path that creates a non-intersecting state cycle (i.e.~each state occurs only once and path ends to the starting point) such that the mean weight of the cycle is maximized (or minimized).  By using the steady state solution $P^{\rm S}$ of the ME model with matrices for time-independent rates ($H$) and displacements ($d$), one can compute the cycle that maximizes the mean velocity of the island within the non-equilibrium currents that arise in non-zero field.  We call these the \emph{dominating transport cycles}.  This cycle may not be the actual most probable cycle, which would be practically impossible to compute for large graphs (see e.g.~\cite{trajektorit}), but merely an approximation.  For additional details, see Sec.~III.A of Ref.~\cite{Kauttonen2}.

In addition to our previous approach \cite{Kauttonen2}, we here propose an optimization of the type
\begin{equation}
\max_{C} \left\{ \frac{\sum_{\expect{j,i} \in C}^{} d_{j,i}}{\sum_{\expect{j,i} \in C}^{} \frac{1}{J_{j,i}}} \right\},
\label{eq:optimointi}
\end{equation}
where $C$ is a cycle on a directed graph, $d_{j,i}$ ($=-d_{i,j}$) is the displacement in the chosen direction and $J_{j,i} = H_{j,i}P_i^{\rm S} - H_{i,j}P_j^{\rm S} > 0$ is the probability current from state $i$ to $j$ (positive because the graph is directed).  The maximization can be turned into minimization depending on which transport direction one wishes to study.  This type of path maximizes the cycle in the ``distance/time'' sense as opposed to ``velocity/edge'' of our previous work in Ref.~\cite{Kauttonen2}, and it should be better suited for situations where the distances are very unequal and transitions with zero net current are not allowed.  Also this kind of optimization is analogous with the way one usually computes the average velocity.

\subsection{The limit of very large fields}

Although neither the MC or ME model cannot provide a satisfactory approximation of the real system at fields far beyond $0.1~{\rm eV}$ for most materials, it if useful for the analysis of the results to also consider what happens for these models in the limit of very large fields.  Because in the ME model there are no such island configurations where parts of the island are connected only via next-nearest neighbors, it is evident that islands can become locked into configurations that require escaping over several large field-induced energy barriers.  Therefore the escape probability of such configurations approaches zero as the field increases.  These configurations are typically called \emph{trap configurations}, and they also appear for other models \cite{Kolomeisky}.  For the axis-directed field this means variations of U-shapes and for the diagonal field, trap configurations are V-shaped.  Examples of such configurations are found in Fig.~\ref{fig:isokentta}(e)-(f) for $N=11$ and $N=12$.  The situation is analogous with reptating polymers in large fields, for which a large field limit for the velocity can be derived for the Rubinstein-Duke model, having the form $v \propto \exp \left( - E N \right)$ \cite{Kolomeisky}.  Similar behavior can also be expected for the ME model.

For the MC model, there are no trap configurations, since all configurations can be escaped from without going against the field.  Therefore the velocity is not expected to decrease even for very large fields.  For the axis-directed field, the island is expected to finally become a rod (i.e.~a single row of atoms).  For the diagonal field, there is no similar equilibrium shape because of the competition between axis-directed atomic transitions.  This results in complex ``zig-zag''-shape islands, where bulk atoms may still remain.  See the supplemental material  \cite{supplementary} for examples of these configurations.  For very large fields, the oscillatory behavior found at small fields \cite{Rusanen} must finally disappear as the islands are strongly deformed and the number of bulk atoms decreases.  Because of discreteness, one may assume that the transition in island geometry is not smooth as a function of the field amplitude.  This is especially true for axis directed fields, where the average width of the island perpendicular to field can be expected to decrease from $\sim\sqrt{N}$ to unity for increasing field amplitude.

\subsection{Pulsed field and electrophoretic ratchet}

We consider two types of time-dependent fields: the pulsed field and the electrophoretic ratchet.  The variation of the field is taken to be discrete, i.e.~with two constant fields varied temporally corresponding to two sets of rates $\Gamma_{f,i}^k$ with $k \in \left\{ 1,2 \right\}$.  The periods of the fields are $\tau_1$ and $\tau_2$ with the total period $\tau = \tau_1 + \tau_2$ and symmetry parameter $x=\tau_1/\tau$.  

For MC computations, the field variation is deterministic (i.e.~the field periods are exact), whereas for the ME computations stochastic Markovian type switching is applied (i.e.~$\tau_1$ and $\tau_2$ are expected values).  These choices allow best possible computational accuracy for both methods, avoiding serious numerical problems arising from the bad statistics of the Monte Carlo simulations or numerical integration of stiff master equation sets.  See supplemental material for further details and a comparison of switching types \cite{supplementary} for the ME model.  This choice also allows comparison between these types of variation.

For the pulsed field we consider measurement angles $\gamma=0^\circ$ (in the direction of the coordinate axis) and $\gamma=45^\circ$ (the diagonal direction).  The field angles are $\pm \alpha$ (for $\gamma=0^\circ$) and $45^\circ \pm \alpha$ (for $\gamma=45^\circ$).  The velocity is always positive when $0^\circ<\alpha<45^\circ$.  Field periods and amplitudes are taken to be identical (i.e.~$x=1/2$ and $E_1=E_2$), so that the average velocity is always in the measurement direction $\gamma$ (see Fig.~\ref{fig:esimerkkikuva1}).  This type of a pulsating field is used in gel electrophoresis to increase the mobility of the DNA samples.

The electrophoretic ratchet, also known as a zero-integrated field, is defined by choosing $\gamma=0^\circ$ with $\alpha$ taking values $0^\circ$ and $180^\circ$.  As the total force affecting the island is $F=2 N E / k_{\rm B} T$, by choosing $\tau_1 F_1 = \tau_2 F_2$ where forces $F_1$ and $F_2$ are in opposite directions, the mean force is always zero (hence the term ``ratchet'').  In the (perfect) linear response regime (i.e.~$v \propto D_{\rm eff} F$), this leads to zero mean velocity.  Beyond that, non-zero velocity is expected.  If one chooses $E_1 > E_2$, the expected velocity based on the single atom case has always a positive sign.  However, for many atom case $N>1$, the sign depends on the model properties and is generally unknown.  The electrophoretic ratchet is therefore a good tool to study and quantify many-particle effects.

The limit $\tau \rightarrow 0$ leads to mean-field-type rates $\Gamma_{f,i}^{\rm mf} = \left( \tau_1 \Gamma_{f,i}^1 + \tau_2 \Gamma_{f,i}^2 \right) / \tau$.  For the electrophoretic ratchet the corresponding mean-field velocity $v_{\rm mf}$ is then always positive due to the exponential function (velocity in the pulsed field is always positive by above definitions).  At the limit $\tau \rightarrow \infty$ one recovers the adiabatic velocity $v_{\rm ad} := \lim_{\tau \rightarrow \infty} \left\{ v(E_1) \tau_1 + v(E_2) \tau_2 \right\} / \tau$ which is simply the combination of two static field velocities.  For an island of a given size in the electrophoretic ratchet the sign of $v_{\rm ad}$ is determined by the shape of $v(E)$ curves, and what happens between the limits of $\tau$ depends on the individual properties of the model.  The most interesting cases to be considered are indeed the adiabatic limit and finite values of $\tau$, especially those that correspond to mean escape times of the different energy barriers.

\section{Results for the static field}
\label{sec:static_field}

By using the master equation (ME) and Monte Carlo (MC) methods and the corresponding models, we have carried out extensive computations for islands with $N<100$ atoms.  Selected MC simulations were also performed for larger islands up to thousand atoms.  All results are computed with the parametrization given for Cu atoms on the Cu(001) surface.  In this study we consider temperatures $T = 400 \dots 1000~{\rm K}$ and field strengths $E=0 \dots 0.25~{\rm eV}$ varying the field direction (the angle $alpha$)  and the measurement direction (the angle $\gamma$). 
To reduce the amount of data shown below, we present detailed results for the ME model (data with better numerical accuracy) and selected results for the MC model (allowing larger islands).

For better comparison between different values of $E$ and $T$ in the time-dependent field, we have re-scaled $\tau$ such that $\tau =1$ always correspond to the largest rate available to the island.  Therefore the value of $\tau$ in the figures is the multiplicity of the fastest rate in system, which is the jump along the terrace in the field direction (i.e.~$\Delta_{f,i} = 0$ and $E_{f,i}=E$ in Eq.\eqref{eq:taajuudet}) and hence depends on both values $E$ and $T$.

\subsection{Case $N=1$}

It is useful first to consider briefly the case of a single adatom (island with $N=1$) with only axis directed nearest-neighbor jumps.  Since single atom has no internal structure, velocity and diffusion are constants as a function of $\tau$.  For single atom without diagonal movements, using Eq.~\eqref{eq:taajuudet} the velocity takes a form
\begin{align*}
v(E,T,\gamma,\alpha) = 2 \exp \left( -E_S / k_b T \right) \left[ \cos (\gamma ) \sinh \left( \frac{E \cos (\alpha)}{k_b T} \right) + \sin (\gamma ) \sinh
   \left( \frac{E \sin (\alpha)}{k_b T} \right) \right],
\end{align*}
where $\gamma$ is the measurement angle, $\alpha$ field angle and $E$ field amplitude (see Fig.~1).  For $\gamma = 45^\circ$ and suitable values for $E$ and $T$, this function has a maximum value with $45^\circ < \alpha < 135^\circ$ (or equivalently $-45^\circ < \alpha < 45^\circ$), i.e.~rotating the field leads to increased current in the diagonal direction.  Approaching the limit $E/T \rightarrow \infty$ the maximum point shifts toward $90^\circ$ (or $0^\circ$).  This is a straightforward non-linear property of the exponential function.  For the same reason, for $\gamma =0^\circ$ the maximum is always found with $\alpha =0^\circ$, hence the current cannot be further increased by rotating the field.  For a single particle, instead of individual magnitudes for $E$ and $T$, only the ratio $E/T$ is important.  This is not the case for islands with $N>1$.

\subsection{Velocity as a function of field}

\begin{figure}
\includegraphics[width=16cm]{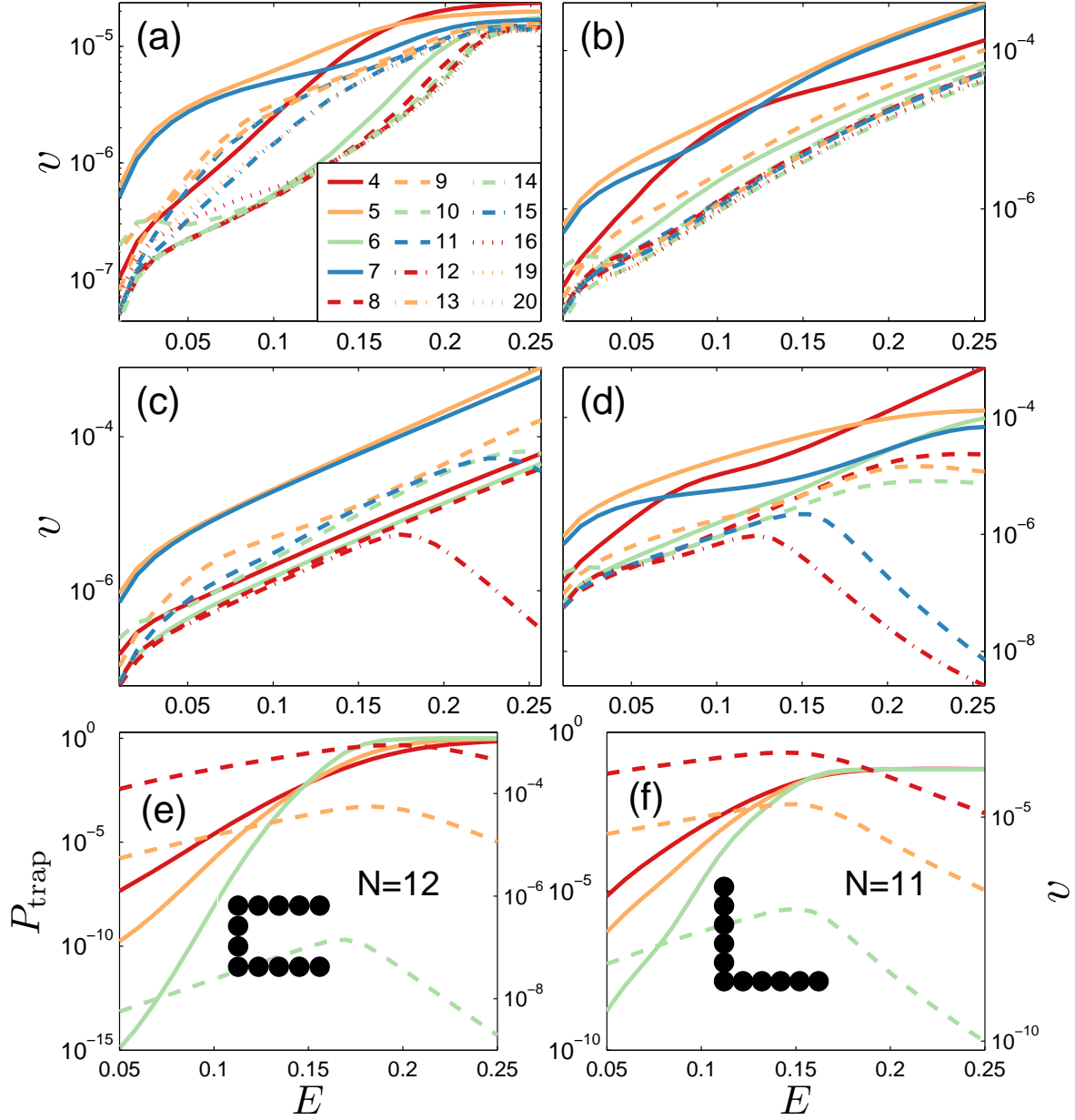}
\caption{(Color online) (a)-(d) Island velocity in the (a and b) MC model $N=4 \dots 20$ and (c and d) ME model $N=4 \dots 12$ at $T=500~{\rm K}$ with (a and c) $\gamma = 0^\circ$ and (b and d) $\gamma = 45^\circ$.  (e)-(f) Velocity (dashed lines) and probabilities $P_{\rm trap}$ (solid lines) of the main trap configurations shown in the inset figures in the ME model for (e) $N=12$ and (f) $N=11$ for temperatures $800~{\rm K}$ (up most), $600~{\rm K}$ (middle) and $400~{\rm K}$ (lowest).  In Figs.~(e) and (f) the vertical axes on the left shows $P_{\rm trap}$ and the axes on the right shows $v$.}
\label{fig:isokentta}
\end{figure}

The velocity as a function of the field is shown for $N=4 \dots 20$ (MC model) and $N=4 \dots 12$ (ME model) in Fig.~\ref{fig:isokentta}, using $\gamma = \alpha$ and $T=500\,{\rm K}$.  There are noticeable differences between ME and MC models especially for $E>0.1~{\rm eV}$ as the velocities begin to decrease for the largest $N>7$ islands in the ME model.  This is caused by the trap configurations as demonstrated in the last row for $N=11$ (f) and $N=12$ (e), where velocities and probabilities of the main trap configurations are shown as a function of $E$ with diagonal (f) and axis directed (e) fields and temperatures $400\,{\rm K}$, $600\,{\rm K}$ and $800\,{\rm K}$.  Note that in the diagonal field there are several energetically equivalent trap configurations for the  $11$-atom island, hence the probability of the main trap configuration does not reach 1.

Within the linear response regime (with field up to $E \sim 0.01\,{\rm eV}$), the velocity is affected by the diffusion coefficient and the ``magic size'' effect strongly affects the velocity for small islands ($N<11$).  In the regime $E>0.1\,{\rm eV}$ the velocity depends strongly on whether $N$ is odd or even and the velocity is significantly larger for odd-$N$ islands.  This effect is stronger for the axis-directed field, where all odd-$N$ islands are faster and curves become ``bunched'' in two distinctive groups with a noticeable gap in between.  At least for smaller islands, this is caused by the fact that the even-$N$ islands easily fall into complete rectangle shapes of width 2 (i.e.~two atom rows).  Escaping this shape requires breaking two nearest-neighbor bonds.  For odd-$N$ islands, such a compact shape is unavailable, hence they have faster transition paths available (this aspect is studied further in Sec.~V).  Around $E \sim 0.1~{\rm eV}$ the velocity behavior clearly changes for all but the smallest islands.  At $E \sim 0.25~{\rm eV}$ the velocity is no longer increasing for the MC model.

The behavior of the island $N=10$ is somewhat special for both the ME model and the MC model (a small-system effect), since at low temperatures the velocity is decreasing for $E=0.02 \dots 0.05\,{\rm eV}$.  At larger fields, the behavior becomes similar to large even-$N$ islands, indicating that 10 atoms is already enough to capture the characteristic behavior of larger islands.

The simulation data (not shown) indicates that in fields $E \sim 0.1~{\rm eV}$ and beyond, the island would be much more likely to break up for diagonal fields when compared to axis-directed fields with the same magnitudes.  The reason for this is that the islands have less atomic bonds on average in a diagonal field, which more easily results into break-up of the island.

\begin{figure}
\includegraphics[width=16cm]{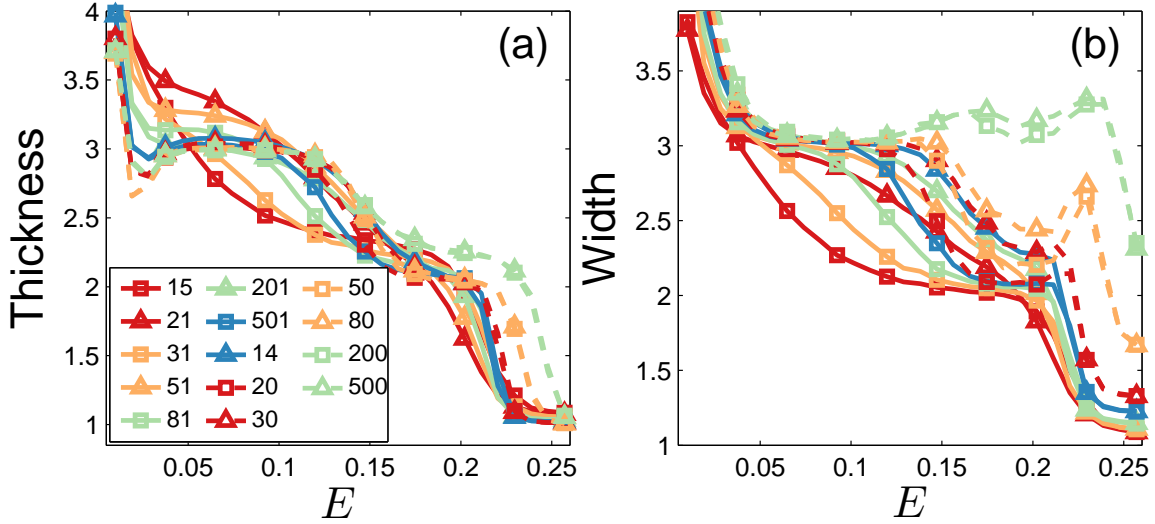}
\caption{(Color online) Change of average geometry of various island sizes at $T=500~{\rm K}$ as a function of field amplitude ($\alpha = \gamma = 0^\circ$) measured by the island (a) average thickness and (b) width perpendicular to the field (see text).  Note that the density of data points is higher than the density of plotting symbols.  The data is for the MC model.}
\label{fig:isokentta_geometriamuutos}
\end{figure}

In Fig.~\ref{fig:isokentta_geometriamuutos} we show the transition of the island geometry as a function of the axis directed field ($\gamma = \alpha = 0^\circ$) for the MC model.  The geometry is characterized by the average thickness and width of the island.  From these quantities, the maximum elongation of the island is measured in both parallel and perpendicular to the field.  The width is given by the perpendicular size and average thickness by the island size divided by the parallel length.

Three distinctive steps corresponding to widths 1, 2 and 3 are seen (i.e.~on average the island consist of 1-3 rows of atoms).  Step 3 becomes visible only for large enough systems ($N \sim 50$), whereas the other two steps are visible for all systems.  For smaller islands ($N<100$) there is a clear even-odd effect for the island size at $E \approx 0.05 \dots 0.20~{\rm eV}$.  The average shape of even-$N$ islands is flatter, which indicates that they are usually found in their tightly-bound rectangle configurations, whereas the sizes of odd-$N$ islands can vary more freely.  The increase of width for large islands in fields $E \sim 0.2\,{\rm eV}$ is caused by configurations where instead of single rod, there are several smaller rods that together occupy adjacent rows and consecutive small rods have single row misplacement in perpendicular axis direction (see Fig.~1 in the supplemental material \cite{supplementary}). Only increasing the field further, a single rod structure with smaller total energy becomes a dominating configuration.  For average thickness, there is also an interesting local minimum at $E \approx 0.02~{\rm eV}$.   Similar behavior can be also found by using other measures, such as the variance of the width (see Fig.~2 in the supplemental material \cite{supplementary}).

\subsection{The effect of measuring and field angles}

\begin{figure}
\includegraphics[width=16.0cm]{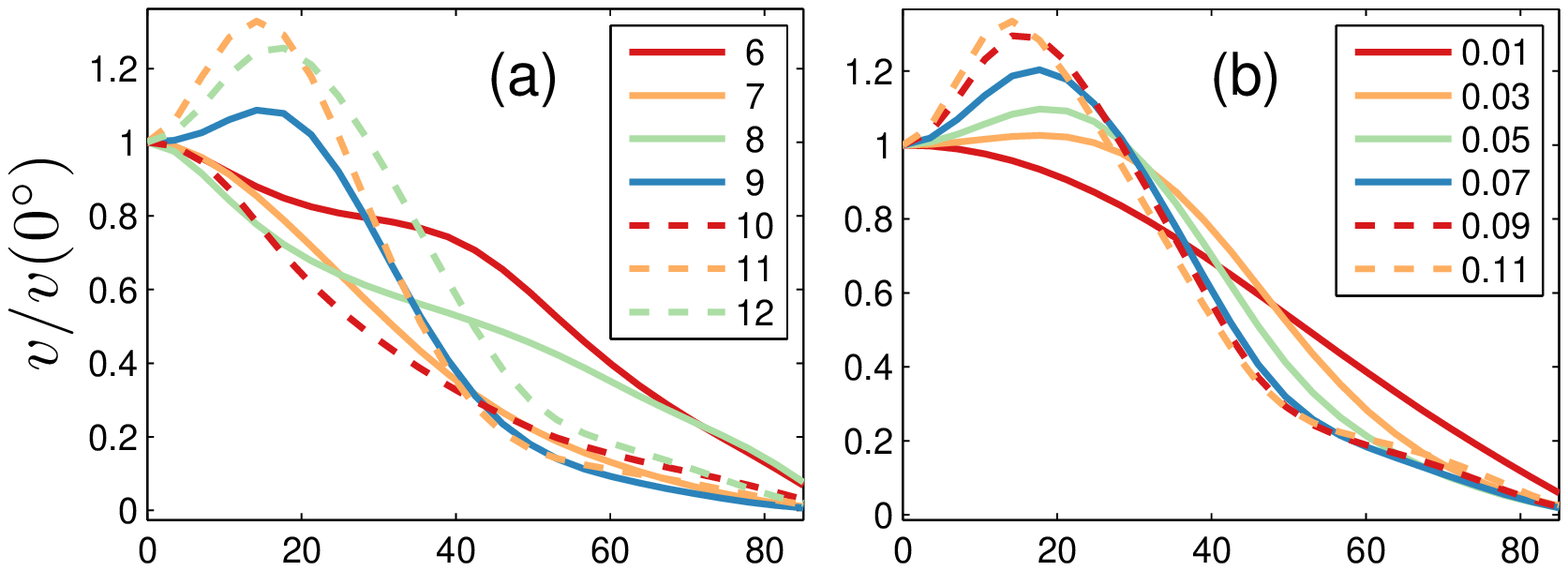}
\includegraphics[width=16.0cm]{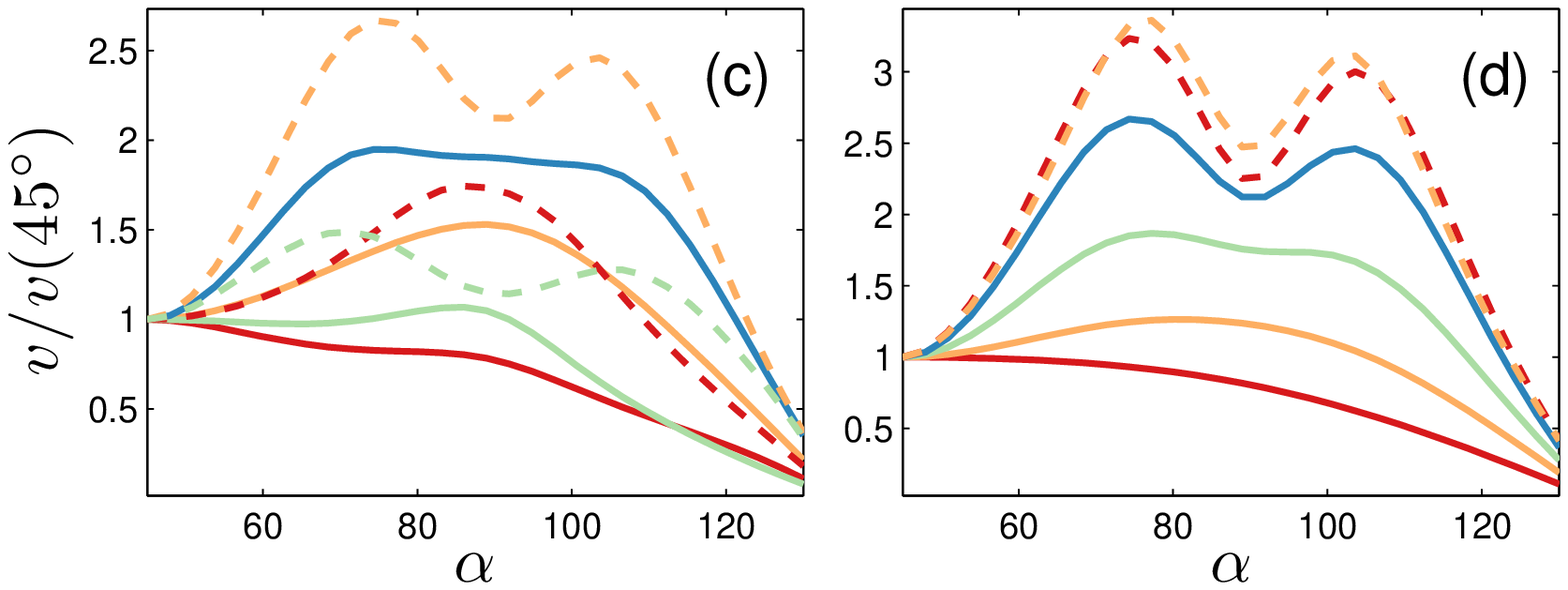}
\caption{(Color online) Velocity increase in directions (a and b) $\gamma=0^\circ$ and (c and d) $\gamma=45^\circ$ as a function of field angles in temperature $T=500~{\rm K}$. (a) Velocity scaled with $v(\alpha=0^\circ)$ for $N=6 \dots 12$ and $E=0.08~{\rm eV}$. (b) Case $N=12$ with several field amplitudes $E=0.01 \dots 0.11~{\rm eV}$. (c) Velocity scaled with $v(\alpha=45^\circ)$ for $N=6 \dots 12$ and $E=0.07~{\rm eV}$. (d) Case $N=11$ with several field amplitudes $E=0.01 \dots 0.11~{\rm eV}$. The data is for the ME model.}
\label{fig:staattinen_kentta}
\end{figure}

In Fig.~\ref{fig:staattinen_kentta} the effect of the field angle $\alpha$ is shown for $N=6 \dots 12$ for the measuring directions $\gamma=0^\circ$ and $\gamma=45^\circ$ computed for the ME model using $E=0.08~{\rm eV}$ and $T=500~{\rm K}$.  To find out the proportional velocity, we scale the results by corresponding $v(\gamma = \alpha)$.

In contrast to the single particle in the case $\gamma = 0^\circ$, the maximum velocity is not always at $\alpha=0^\circ$, but can indeed have a value between $0^\circ<\alpha<90^\circ$ depending on the island size.  Increasing $E$ and decreasing $T$ leads to increased velocity, however the field $E$ must be large enough for non-linear effect to appear.  The maximum velocity is found with $\alpha = 15 \dots 25^\circ$ for islands over $10$ atoms.  Rotating the field slightly (i.e.~increasing $|\gamma - \alpha|$) creates a small field component in $y$ direction.  This decreases the barrier for corner rounding process for the other side of the island, which leads to increased velocity.

For the measuring direction $\gamma=45^\circ$ the maximum velocity is found for $45^\circ <\alpha < 90^\circ$.  This is expected from the single particle case.  However, there are two local maxima for islands of size $N>10$ located both sides of the angle $\alpha = 90^\circ$, creating a small deviation of approximately $15^\circ$ from angle $90^\circ$.  The global maximum is found around $\alpha \approx 70^\circ$ and the second one around $105^\circ$.  For the smallest islands $N<7$, no increase is found.  As demonstrated for the case $N=11$, a two-maximum structure appears when the field gets strong enough.  Transport in the diagonal direction is generally more difficult compared to the axis direction because of the absence of stable rectangular configurations.  By rotating the field, rectangular shapes become stable and the velocity increase occurs for a much larger range of field angles than in the case of $\gamma = 0^\circ$.

\begin{figure}
\includegraphics[width=17.0cm]{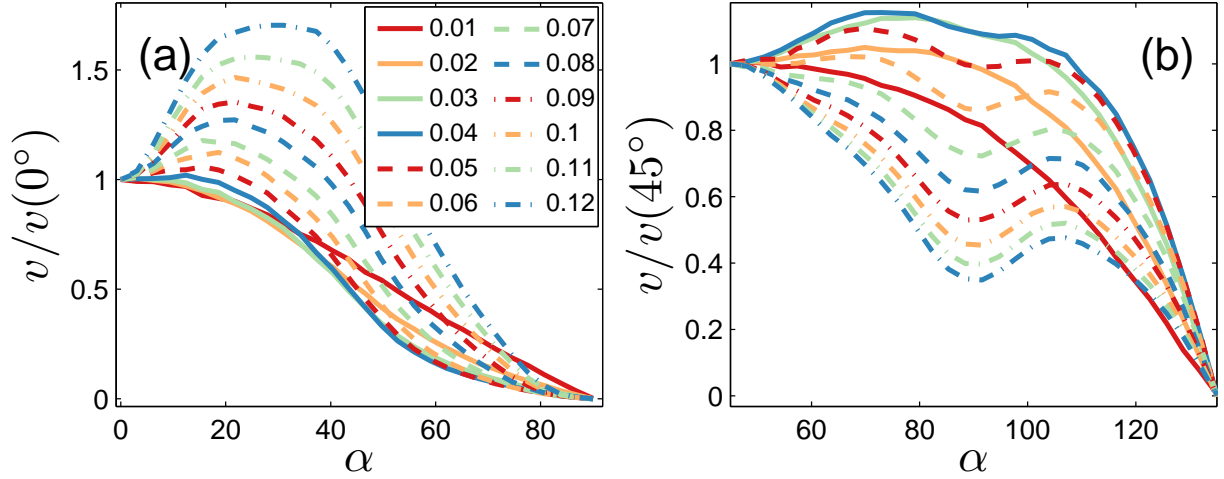}
\caption{(Color online) Velocity increase of $N=20$ with $T=600~{\rm K}$ as function of $\alpha$ for several field amplitudes $E=0.01 \dots 0.12~{\rm eV}$. (a) Case $\gamma=0^\circ$. (b) Case $\gamma=45^\circ$.  The data is for the MC model.}
\label{fig:staattinen_kentta_N20}
\end{figure}

The findings above can be also verified for the MC model.  In Fig.~\ref{fig:staattinen_kentta_N20} data for $N=20$ is shown.  For $\gamma=0^\circ$, the maximum is found around $\alpha =15 \dots 25^\circ$ for smaller fields but shifts up to $35^\circ$ for large fields ($E>0.1~{\rm eV}$).  This shift is not present for the ME model.  The MC model also creates a strong odd-even $N$ effect that is not present for the ME model. For fields above $0.1~{\rm eV}$, double maxima structure appears for large islands $N \sim 20$ and beyond (see supplemental material \cite{supplementary}).  For $\gamma = 45^\circ$ the results are more consistent with the ME model and the two-maxima structure is visible with the global maxima found for angles for $\alpha  \approx 70^\circ$.  In contrast with the case $\gamma=0^\circ$, the velocity increase eventually disappears for very large fields.

\begin{figure}
\includegraphics[width=17.0cm]{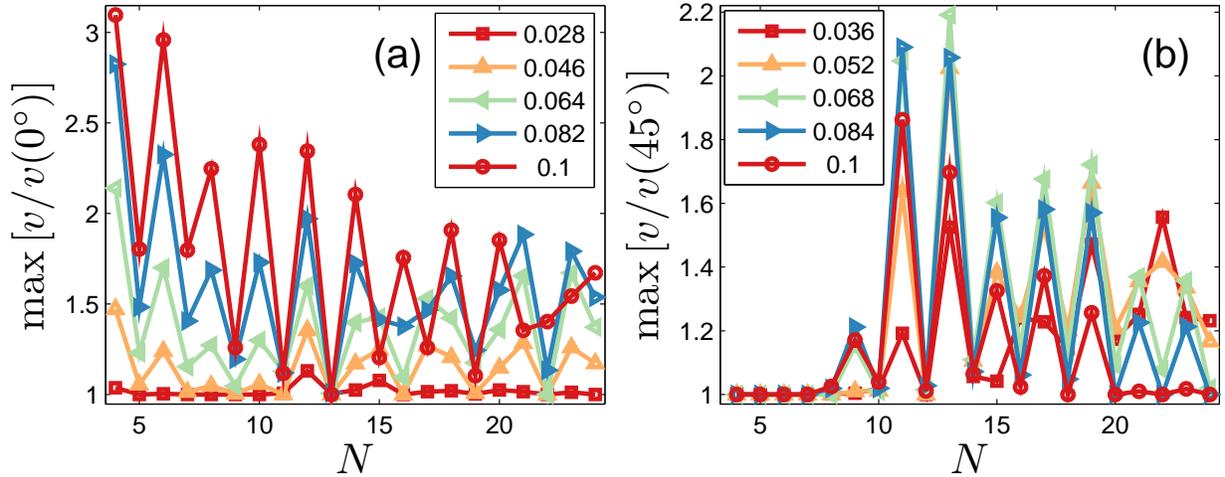}
\caption{(Color online) Maximum velocity increase for $N=4 \dots 24$ compared to (a) $v(0^\circ)$ and (b) $v(45^\circ)$ for several field amplitudes $E=0.028 \dots 0.1~{\rm eV}$ in $T=500~{\rm K}$.  The data is for the MC model.}
\label{fig:staattinen_kentta_maksimit}
\end{figure}

To further clarify the odd-even effect and the amount of increase for the velocity, the maximum increase of the velocity is plotted in Fig.~\ref{fig:staattinen_kentta_maksimit} for islands $N=4 \dots 24$ with several field amplitudes using the MC model.  The odd-even effect is strong for $N<15$ and only for larger islands deviations from this rule begin to appear.  The results for the MC model show that the rotated field favors even-$N$ islands in the case $\gamma = 0^\circ$ and odd-$N$ islands in the case $45^\circ$.  The optimal angles for odd-$N$ islands are smaller than for the even-$N$ islands in the case $\gamma = 0^\circ$, whereas for the case $\gamma = 45^\circ$ the behavior is just the opposite.

The key conclusion here is that beyond the linear-response regime, the velocity depends strongly on the measurement and field directions and the velocity can be significantly increased by setting a small $15^\circ \dots 25^\circ$ difference between field and measurement directions.  This can be exploited by using a time-dependent field.  For the case $\gamma=45^\circ$ velocity increase can be expected based on the single atom case, however the optimal field angle for islands is not $\alpha \approx 90^\circ$, but has a double maxima structure with optimal angles around $70^\circ$ and $105^\circ$.  This deviation from the single atom case results from the corner process. By introducing a small non-axis-directed field component, going around the corner is made easier.  For the same reason, velocity increase is also present in the case $\gamma=0^\circ$, where the maximum velocity is found with field angles $\alpha \approx 20^\circ$. It is also found that there is a strong odd-even island-size effect affecting the amount of velocity increase and also the values of optimal $\alpha$.  For the case $\gamma=0^\circ$, even-$N$ islands become significantly faster and for the case $\gamma=45^\circ$ the behavior is just the opposite.  The results differ between the ME and MC models, especially for the case $\gamma=0^\circ$, since the ME model does not reproduce the odd-even effect or the increase for the smallest islands.  This indicates that the configurations with only diagonal bonds between parts of the island, present only in the MC model, become important in this particular situation.

\subsection{Effective energy barriers}

An effective energy barrier can be found via Arrhenius plots $\ln ( D_{\rm eff} )$ or $\ln ( v )$ vs. $1 / k_{\rm B} T$.  If the effective barrier is constant for a large temperature interval, it means that the transport process is similar in that region and a data collapse is possible.  In the previous studies concerning equilibrium and very small fields, it has been found that the effective barrier is around $0.75~{\rm eV}$ for large islands $N > 10$ \cite{Merikoski2,Karimi1,Skodje,Mattsson} and varies between $0.5 \dots 0.8~{\rm eV}$ for the smallest islands \cite{pieni_arrhenius}.  With nearest-neighbor count energetics, this is roughly equivalent to transitions that break two nearest-neighbor bonds.  The effective barrier is typically lower for small islands and in higher temperatures \cite{Mattsson,Voter}.  We used the ME model to compute accurately the temperature dependent effective barriers for small islands for several field amplitudes.  Because of a large statistical error, a similar study would be complicated by using only simulation data.  Here we set $\alpha = \gamma$.  Because of the computational difficulties in low temperatures (especially for $D_{\rm eff}$), we show only those values that remain reliable and omit the results for the lowest temperatures where data becomes noisy.

\begin{figure}
\includegraphics[width=16.0cm]{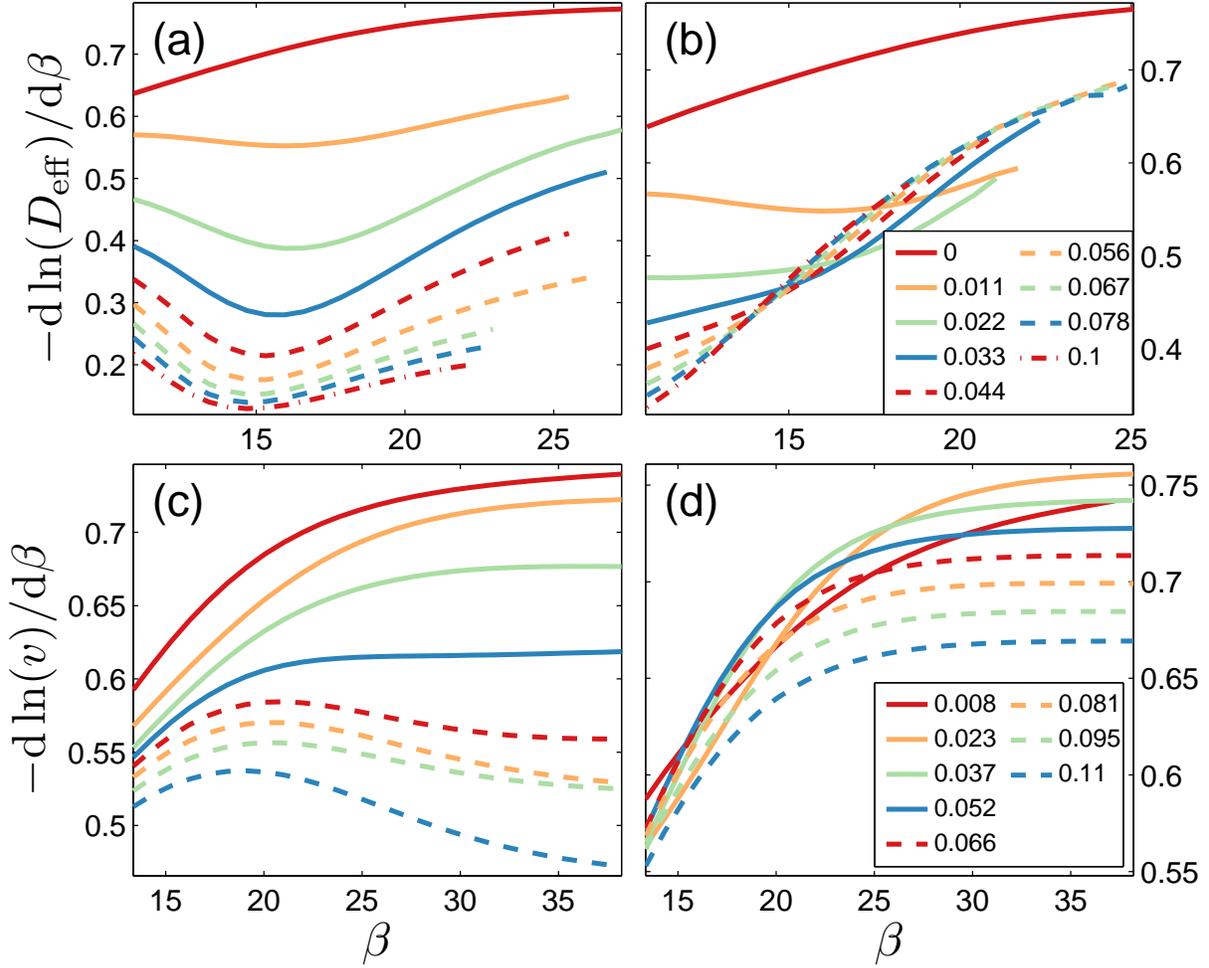}
\caption{(Color online) Running slope of the Arrhenius curves computed using (a and b) $D_{\rm eff}$ and (c and d) $v$ for (a and c) $N=11$ and (b and d) $N=12$ with axis directed field (i.e.~$\gamma = \alpha = 0^\circ$) and amplitudes $E=0 \dots 0.1~{\rm eV}$.}  
\label{fig:arrhenius_diff}
\end{figure}

\begin{figure}
\includegraphics[width=16.0cm]{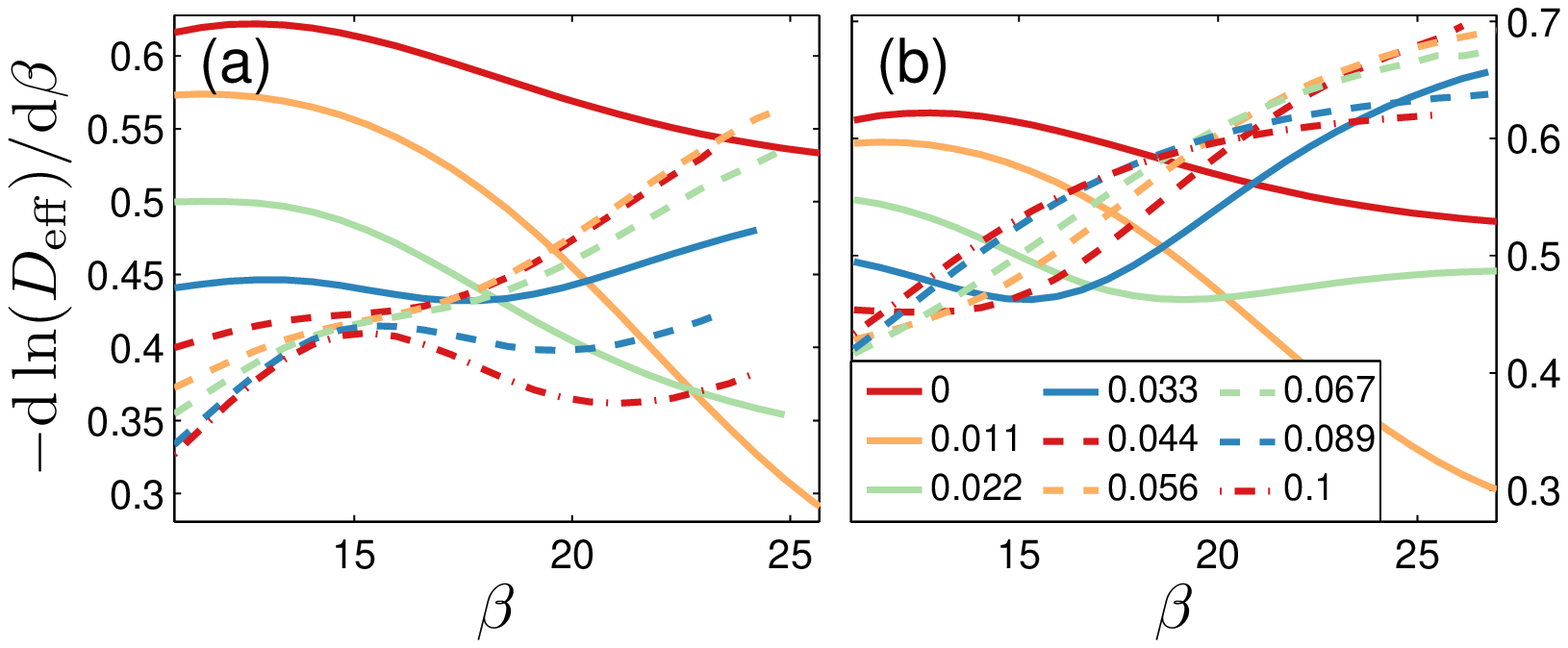}
\caption{(Color online) Running slope of the Arrhenius curves for a special case $N=10$ computed using $D_{\rm eff}$ with (a) $\gamma = \alpha = 0^\circ$ and (b) $\gamma = \alpha = 45^\circ$ for several field amplitudes $E=0 \dots 0.1~{\rm eV}$.}
\label{fig:arrhenius_diff_N10}
\end{figure}

In Fig.~\ref{fig:arrhenius_diff} the running slope of the Arrhenius curve or the effective activation barrier is shown for several field amplitudes for $N=11$ (a and c) and $12$ (b and d) using both $D_{\rm eff}$ (a and b) and $v$ (c and d) in direction $\gamma = 0^\circ$.  In zero field, an effective barrier around $0.7~{\rm eV}$ is found with only a minor temperature dependence.  However, as the field gets stronger, the effective barrier depends strongly on the temperature.  At temperatures around $700 \dots 800~{\rm K}$ a distinctive local minimum is found for $N=11$ using $D_{\rm eff}$, which indicates some type of change in the diffusive property of the island transport.  For $v$, there is a local maximum instead of a minimum.  A strong even-odd effect is visible. In low temperatures, the field has only a minor effect on the effective barrier for even islands, whereas the effect is large for odd islands.  The spread for the effective barriers is much smaller for the scaling of $v$ when compared to that computed using $D_{\rm eff}$, otherwise the behavior is similar.  The behavior for $\gamma =45^\circ$ is found to be very similar and is not shown here.  Since islands $N=11$ and $N=12$ already have characteristics of large islands (see \cite{Voter} and Sec.~VI), we expect similar behavior to be observed also for somewhat larger islands.  We note however that results for $N=10$ were found atypical from other small islands $N=8 \dots 12$, because of the different temperature dependency and also for the large differences between axis and diagonal fields.  Results for $N=10$ are shown in in Fig.~\ref{fig:arrhenius_diff_N10}.  The temperature dependence of the effective barrier is strongly affected by the field and also the differences between $\gamma =0^\circ$ and  $\gamma =45^\circ$ are large.

The results above indicate that there is no well-defined effective energy barrier in the presence of field.  In addition to the field amplitude, the effective barrier depends strongly on the temperature, especially for temperatures above $500~{\rm K}$.  In the general case, velocities and effective diffusion coefficients of small islands do not follow any simple scaling laws such as those withing the linear response regime \cite{Karimi1}.  Also the effective barriers for diffusion and the velocity differ.

\subsection{The leading relaxation constant}

By computing the second highest eigenvalue of the stochastic generator $H$ (the highest one being zero), the leading relaxation time ($\Phi$) can be found as the inverse of the eigenvalue.  This relaxation time is the property of the linear set of equations and is independent of the initial state, hence it is not directly related to the relaxation time found in experiments or simulations where one usually measures the relaxation of some macroscopic observables, such as the shape and the size of the islands \cite{relaksaatio}.  Instead, it has a large effect on finding a numerical solution of the steady state for both time-dependent and time-independent types of potential using numerical integration or iterative eigenvalue solvers.  As the leading relaxation time increases (i.e.~the second eigenvalue approaches zero), the search for the steady state becomes more time-consuming and error-prone.  The expected result is that the relaxation time decreases monotonously as the field gets stronger.  However, because of the trap configurations of the ME model, the relaxation times must eventually become rapidly increasing for very large fields ($E>0.1~{\rm eV}$).  We are aware of only one previous study where the relaxation constant and few other eigenvalues were computed directly, but the vacancy islands using a discretized continuous-space model \cite{spektri}.

\begin{figure}
\includegraphics[width=11.0cm]{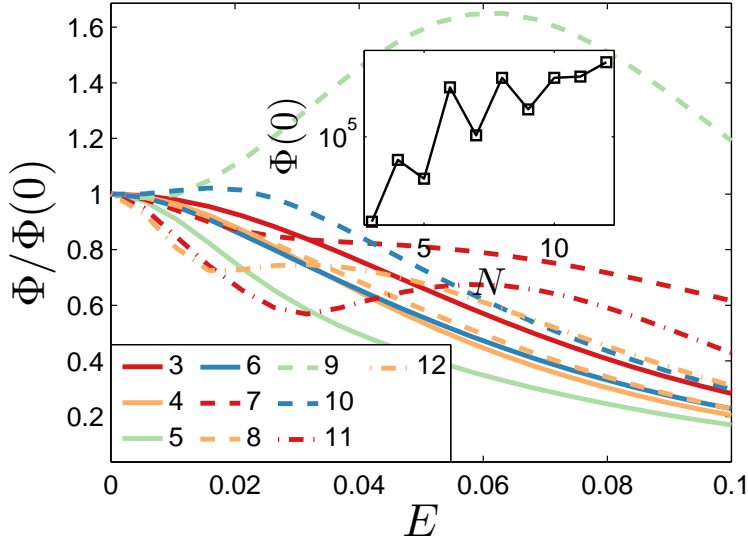}
\caption{(Color online) Leading relaxation constant $\Phi$ as a function of field amplitude for $N=3 \dots 12$ re-scaled with zero field values (shown in inset) with $T=600~{\rm K}$ and $\gamma=\alpha=0^\circ$.}
\label{fig:staattinen_kentta_relaksaatio}
\end{figure}

In Fig.~\ref{fig:staattinen_kentta_relaksaatio} we show the absolute relaxation times $\Phi$ in zero field (inset figure) and as a function of the axis directed field for $N=3 \dots 12$ in $T=600~{\rm K}$.  Within the linear response regime, the relaxation times are indeed decreasing for all $N$.  However, for $N=9 \dots 12$ there exists local maxima with $E= 0.02 \dots 0.06~{\rm eV}$.  This effect is not caused by intersections with other eigenvalues, it is a genuine property of the second eigenvalue (as identified at $E=0$).  For $N=9$ the phenomenon is strongest.  The locations and heights of the maxima are slightly shifted by changing the temperature.  For fields beyond $E>0.1~{\rm eV}$ we can confirm that that relaxation times for $N>7$ become rapidly increasing, however the computations become cumbersome and the data is noisy because of the instability of solving eigenvalues of highly non-symmetric matrices.

The maximum seems to appear shortly after the field amplitude reaches the non-linear regime.  The location of the maximum is around $0.06~{\rm eV}$ for island sizes 9 and 11 and around $0.03~{\rm eV}$ for island sizes 10 and 12.  Also, this effect seems to become weaker as the island size increases from $N=9$ to $N=12$. The increased relaxation time does not have an evident correlation with the transport properties considered in Sections III-IV.  Although for $N=10$ and $N=12$ a slight correlation can be seen with Figs.~\ref{fig:arrhenius_diff}(b) and (d) and also \ref{fig:arrhenius_diff_N10}, where the effective barrier turns from decreasing into increasing around $0.03~{\rm eV}$, this cannot be directly related to any microscopic processes.

\section{Results for the time-dependent field}

\subsection{Pulsed field}

From the static-field results, we expect that the velocity can be increased by rotating the field.  This can be utilized by using a pulsating field in such it causes transport in the direction specified by $\gamma$, such that the field period $\tau$ is larger than the relaxation time of the island (the adiabatic limit).  In the following we study the behavior for small values of $\tau$. As before, we re-scale the velocities with the corresponding velocity without rotation (i.e.~$v(\alpha = \gamma$)).

\begin{figure}
\includegraphics[width=10.0cm]{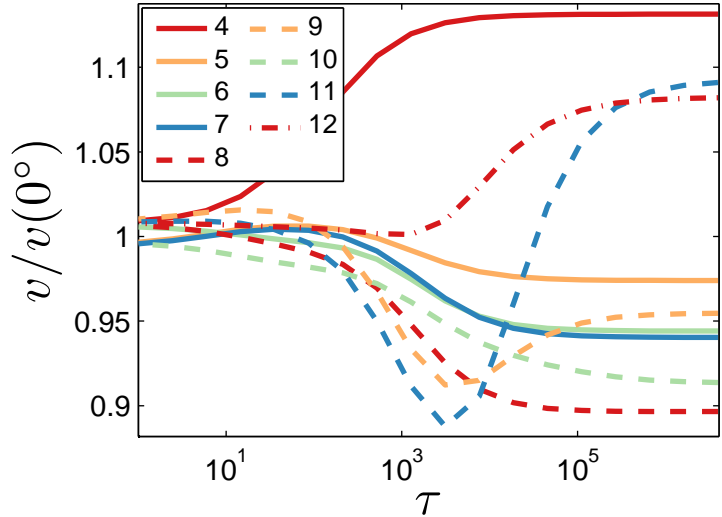}
\caption{(Color online) Velocity increase in the pulsed field for $N=4 \dots 12$, $T=500~{\rm K}$, $E=0.06~{\rm eV}$ with $\gamma = 0^\circ$ and $\alpha = \pm 10^\circ$.  The data is for the ME model.}
\label{fig:pulsed1_0}
\end{figure}

In Fig.~\ref{fig:pulsed1_0} we show velocities for $N=4 \dots 12$ with $T=500~{\rm K}$ and $\alpha = \pm 10^\circ$ compared to the static field velocity in the direction $\alpha = \gamma = 0^\circ$.  For all but $N=10$, the velocity can be slightly increased for small $\tau$.  Odd-$N$ islands have a distinctive local maximum around $\tau = 10^1 \dots 10^2$, but the largest increase occurs typically at the adiabatic limit.  For larger islands $N>10$, minimum velocity is found with $\tau \sim 10^3 \dots 10^4$.

\begin{figure}
\includegraphics[width=17.0cm]{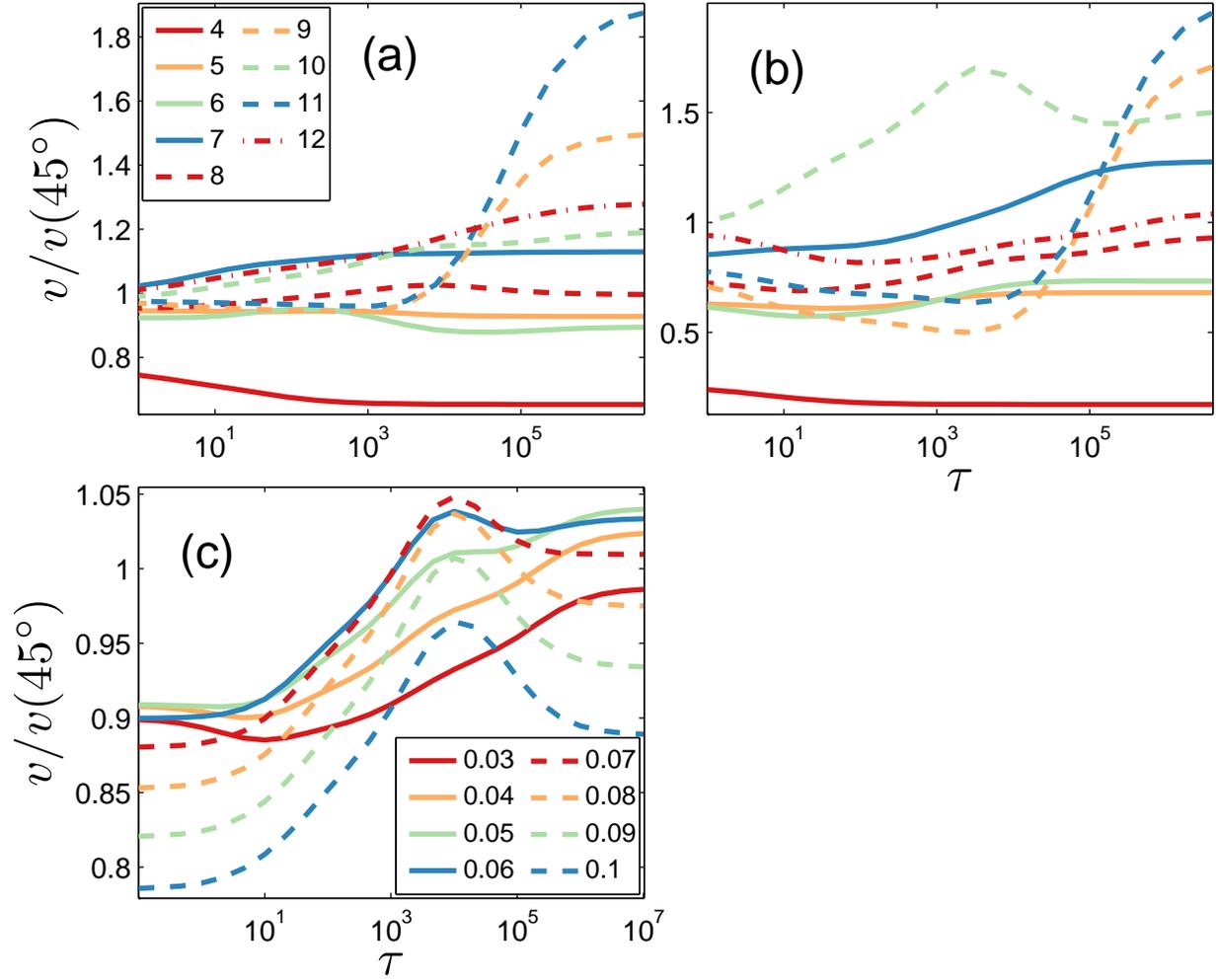}
\caption{(Color online) Velocity increase in the pulsed field for several small islands with $\gamma = 45^\circ$ and $T=500~{\rm K}$.  (a)-(b) $N=4 \dots 12$, $T=500~{\rm K}$, $E=0.06~{\rm eV}$ with (a) $\alpha = 45 \pm 20^\circ$ and (b) $\alpha = 45 \pm 50^\circ$. (c) Distinctive behavior of $N=8$ with $E=0.03 \dots 0.1~{\rm eV}$ and $\alpha = 45 \pm 30^\circ$.}
\label{fig:pulsed1_45}
\end{figure}

\begin{figure}
\includegraphics[width=16.0cm]{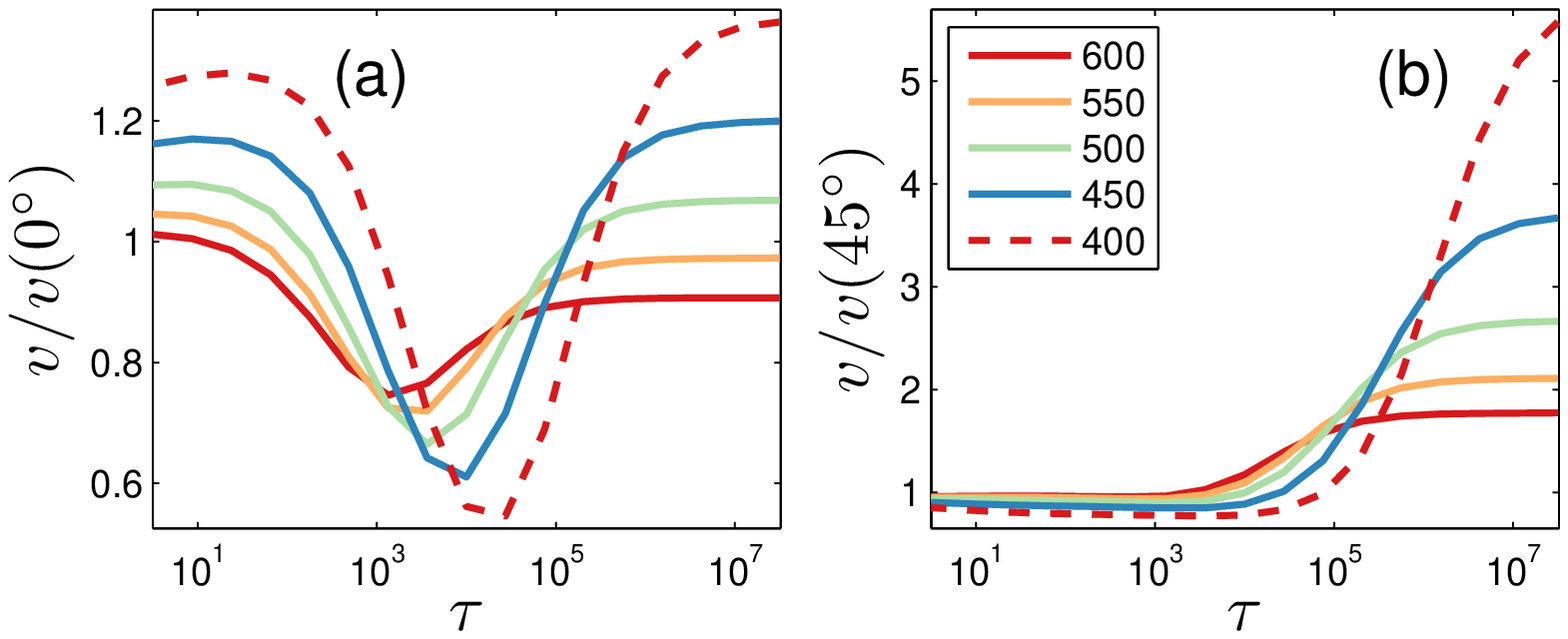}
\caption{(Color online) Velocity increase in the pulsed field for $N=11$ and $E=0.07~{\rm eV}$ for various temperatures $T=400 \dots 600$ with (a) $\gamma=0^\circ$ and $\alpha= \pm 22^\circ$ and (b) $\gamma=45^\circ$ and $\alpha=45 \pm 30^\circ$. The data is for the ME model.}
\label{fig:pulsed2}
\end{figure}

In Fig.~\ref{fig:pulsed1_45} we show similar results for $\gamma = 45^\circ$.  Because of the double-maximum structure, we show results for both  $\alpha = 45 \pm 20^\circ$ and $45 \pm 50^\circ$ keeping other parameters the same as before.  Again, the large $\tau$ limit yields the largest velocity for almost all islands.  For $N \in \left\{ 8,10,12 \right\}$ small local maxima occurs around $\tau = 10^4 \dots 10^5$ for suitable parameters.  This is demonstrated in detail in Fig.~\ref{fig:pulsed1_45}(c) for $N=8$ for several field amplitudes.  For $N>10$, the minimum is again created around $\tau \sim 10^4$. The effect of temperature on the velocity increase is demonstrated in Fig.~\ref{fig:pulsed2} for $N=11$ for both cases of $\gamma$.  Lowering the temperature results in a larger relative velocity increase and also makes the dependence on $\tau$ stronger.

\begin{figure}
\includegraphics[width=16.0cm]{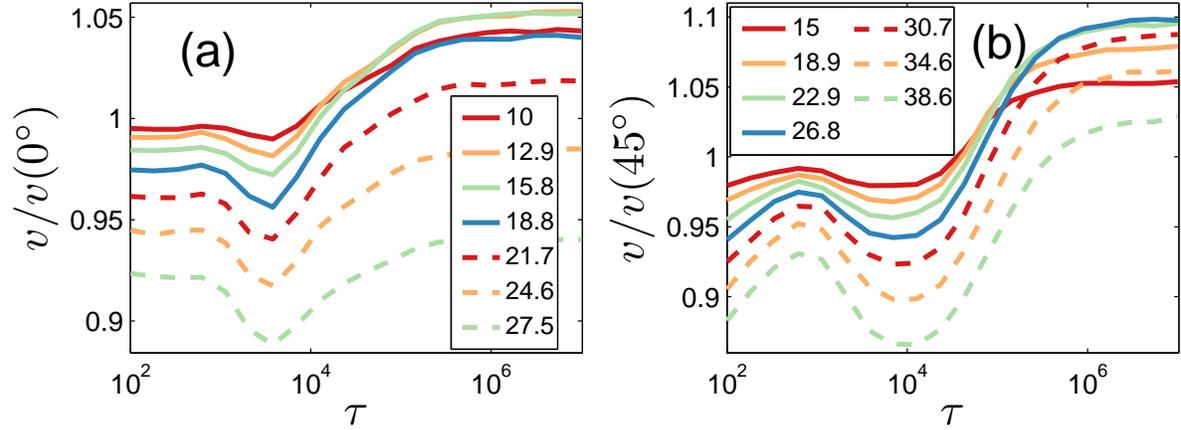}
\caption{(Color online) Velocity increase in the pulsed field with several pairs of the field angle $\alpha$ (values $|\alpha - \gamma|$ are shown in the figure) for $N=20$ and $T=600~{\rm K}$ with field amplitude $E=0.05~{\rm eV}$ for (a) $\gamma = 0^\circ$ and (b) $\gamma = 45^\circ$. The data is for the MC model.}
\label{fig:pulsed5}
\end{figure}

Similar results are also found for larger islands using the MC model.  Although there is some structure (such as local maxima) for small $\tau$ values for large $N$, there is no longer any noticeable increase for the velocity for small $\tau$'s.  Increase is found only at the adiabatic limit.  In Fig.~\ref{fig:pulsed5} we show the velocity for $N=20$ with several values of $\alpha$ for $\gamma =0^\circ$ and $\gamma =45^\circ$.

The results indicate that the steady state velocity of the islands in the pulsed field depends strongly on the period $\tau$ for small islands, but the dependency becomes weaker for large islands $N>10$.  For large islands a significant increase of the velocity is found only at the adiabatic limit (large $\tau$) for both $\gamma = 0^\circ$ and $\gamma = 45^\circ$.  A small $\tau$ tends to increase the velocity of small islands.  For $\tau \sim 10^4$, there is a velocity minimum for all large and also many small islands, indicating the sensitivity to this specific period.  Alternatively, the same period produces a maximum velocity for some small even-$N$ islands.  As already found for the static field case (Sec.~III.D), lowering the temperature significantly increases the sensitivity of velocity to $\tau$.

\subsection{Electrophoretic ratchet}

\begin{figure}
\includegraphics[width=9.0cm]{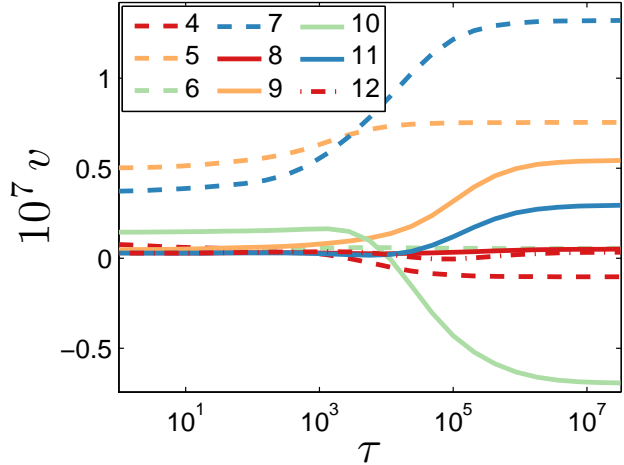}
\caption{(Color online) Velocity in the electrophoretic ratchet with $N=4 \dots 12$ and $x=1/4$ as a function of $\tau$ with $T=500~{\rm K}$ and $E_1=0.03~{\rm eV}$. The data is for the ME model.}
\label{fig:kenttaraikka1}
\end{figure}

\begin{figure}
\includegraphics[width=16.0cm]{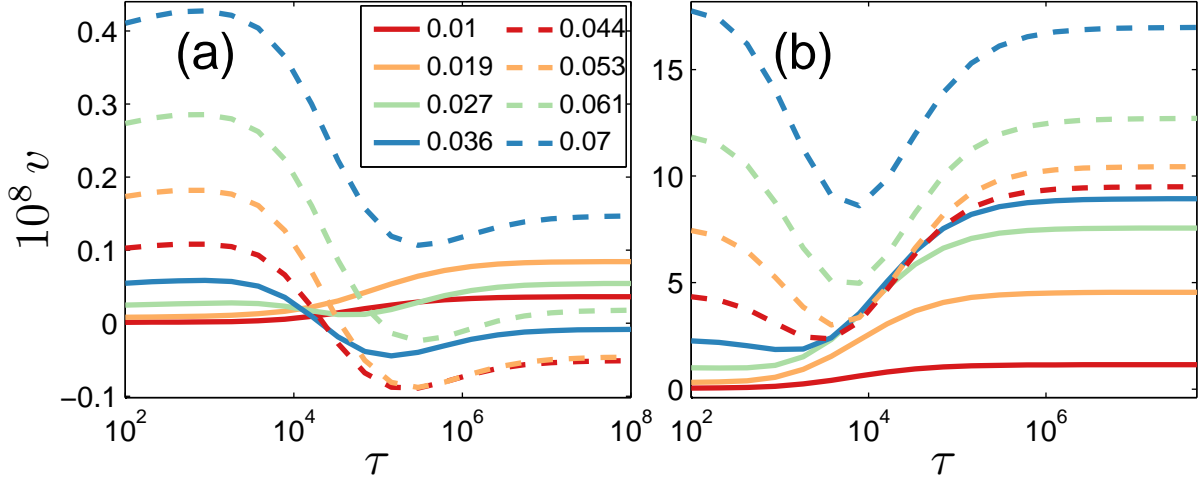}
\caption{(Color online) Velocity in the electrophoretic ratchet with $N=12$ and $x=1/4$ as a function of $\tau$ for several field amplitudes $E=0.01 \dots 0.07~{\rm eV}$ with (a) $T=500~{\rm K}$ and (b) $T=700~{\rm K}$. The data is for the ME model.}
\label{fig:kenttaraikka2}
\end{figure}

In Fig.~\ref{fig:kenttaraikka1} we show the velocity of islands $N=4 \dots 12$ in the electrophoretic ratchet as a function of $\tau$ for the ME model using $T=500~{\rm K}$, $x=1/4$ and $E_1=0.03~{\rm eV}$ (i.e.~$E_2 = 0.03x / (1-x) = 0.01~{\rm eV}$). For $N \in \left\{ 4,10,12 \right\}$, there is a current inversion  for all other $N$ the velocity remains positive.  For $N=4$ and $N=10$, the current inversion is of the adiabatic type (i.e.~results from the static field drifts), for $N=12$ the inversion is of the time-dependent type.  The velocity for $N=12$ is shown in more detail in Fig.~\ref{fig:kenttaraikka2} with temperature $T=500~{\rm K}$ and $T=700~{\rm K}$ and several field amplitudes.  The temperature and the field amplitude have a very large effect on the velocity in the electrophoretic ratchet.  The current inversion easily disappears for increasing the temperature or the field amplitude.

\begin{figure}
\includegraphics[width=17.0cm]{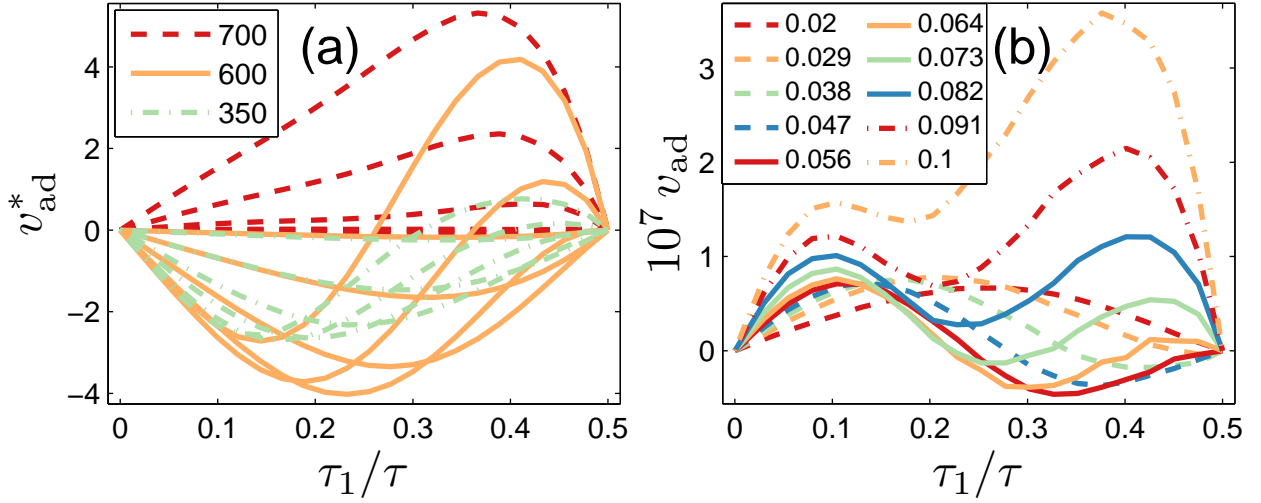}
\caption{(Color online) Adiabatic velocity in the electrophoretic ratchet as a function of ratio $x=\tau_1/\tau$.  (a) $N=10$ with several field amplitudes $E_1=0.01 \dots 0.074~{\rm eV}$ and temperatures $700~{\rm K}$, $600~{\rm K}$ and $350~{\rm K}$ for the ME model.  Arbitrary scaling is used in each temperature for better comparison.  (b) $N=20$ with various field amplitudes $E=0.02 \dots 0.1~{\rm eV}$ in temperature $600~{\rm K}$ for the MC model.}
\label{fig:kenttaraikka4}
\end{figure}

As discussed in Sec.~II.D, the adiabatic limit for the velocity is determined by the velocity in a static field.  The effect of the symmetry parameter $x=\tau_1/\tau$ with $T$ and $E_1$ is demonstrated in Fig.~\ref{fig:kenttaraikka4}, where the adiabatic velocity is computed for $N=10$ and $N=20$.  The adiabatic velocity inversion is typical for a $10$-atom island, whereas the behavior show for $N=20$ is typical for all other large islands.  Choosing the temperature and field strength properly, the adiabatic velocity can be negative or positive and even change sign as a function of $x$.  However, this effect becomes weaker as the island size increases and negative velocities can be achieved for large islands only by fine-tuning field and temperature.  For $N>100$ the negative adiabatic velocity becomes essentially nonexistent and only the positive adiabatic velocity is expected.

\begin{figure}
\includegraphics[width=17.0cm]{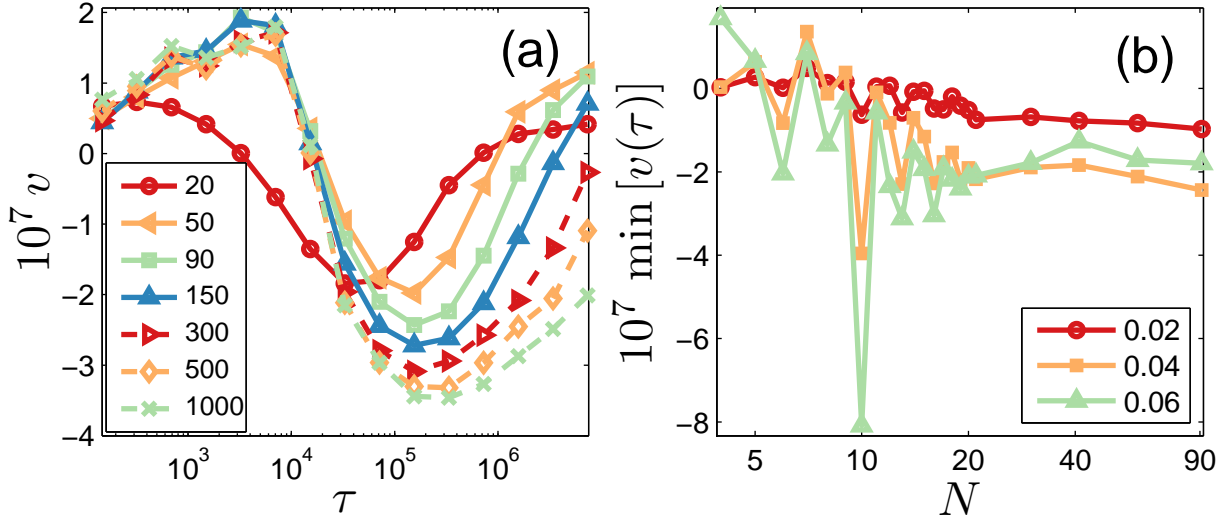}
\caption{(Color online) Velocity in the electrophoretic ratchet with $T=600~{\rm K}$ and $x=1/4$ for the MC model. (a) Velocity for the large islands as a function of $\tau$ for $E_1=0.04~{\rm eV}$. (b) Minimum velocities for several islands $N=4 \dots 91$ for $E_1 = 0.02 \dots 0.06~{\rm eV}$.}
\label{fig:kenttaraikka5}
\end{figure}

The results for the MC model are similar.  There are indeed deep minima for $\tau = 10^4 \dots 10^6$ for large islands which creates a current inversion.  This is shown in Fig.~\ref{fig:kenttaraikka5} for several islands.  For large islands, two local maxima appear at $\tau = 10^3 \dots 10^4$.  Since the current inversion occurs typically only at finite values of $\tau$, it is indeed caused by the interaction between time-dependent field and atoms.  For small islands $N<15$, there is a strong odd-even island size dependency which eventually disappears for larger islands.  For the electrophoretic ratchet this odd-even effect becomes important already for much smaller fields than in the case of a static field.  This is because the ratcheting mechanism with an alternating field direction tends to force islands into thin rectangle shapes.

We conclude that the $\tau$-dependency in the electrophoretic ratchet is found to be much stronger than in the case of the pulsed field.  With suitable field periods $\tau$ around $10^4 \dots 10^6$, the velocity inversion occurs for all large islands ($N>10$) and also for smaller even-$N$ islands.  Especially for the smaller islands $N<20$, the inversion depends strongly on the temperature and field strength, disappearing at large temperatures.  When compared to the velocity increase for a pulsed field, inversion phenomena are observed already with very small field amplitudes near the linear response regime ($E \sim 0.01~{\rm eV}$).  An electrophoretic ratchet have been previously studied within the context of reptating polymers where a similar type of current inversion was found as a function of polymer size \cite{Kulakowski}.

\section{Transition paths}

In this section, we present typical transport mechanisms for small islands using the ME model.  We found that the dominating transport cycles (see Sec.~II.B.2) for time-dependent fields are usually the same as those for the static field, especially for the electrophoretic ratchet where two directions of motion are competing.  Also the results for the $\gamma=0^\circ$ and $\gamma=45^\circ$ are qualitatively similar (the "zig-zag" configurations appear only for islands much larger than $N=12$).  For time-dependent fields, dominating cycles differ from the static field case only for small values of $\tau$, for which the islands have no time to go through a full static-field-type cycles before the potential is changed.  Therefore the dominating transporting cycles cannot be used to explain the velocity increase by a pulsed field in large $\tau$ limit or velocity inversion for the electrophoretic ratchet with $\tau=10^4 \dots 10^6$.  In the following we set $T=600~{\rm K}$ and try various field amplitudes $E$ and report a few optimal cycles given by Eq.~\eqref{eq:optimointi}.  However, if was found that the results are often the same as those that were computed using the different types of optimal cycles proposed in Ref.~\cite{Kauttonen2}.

\begin{figure}
\includegraphics[width=7.5cm]{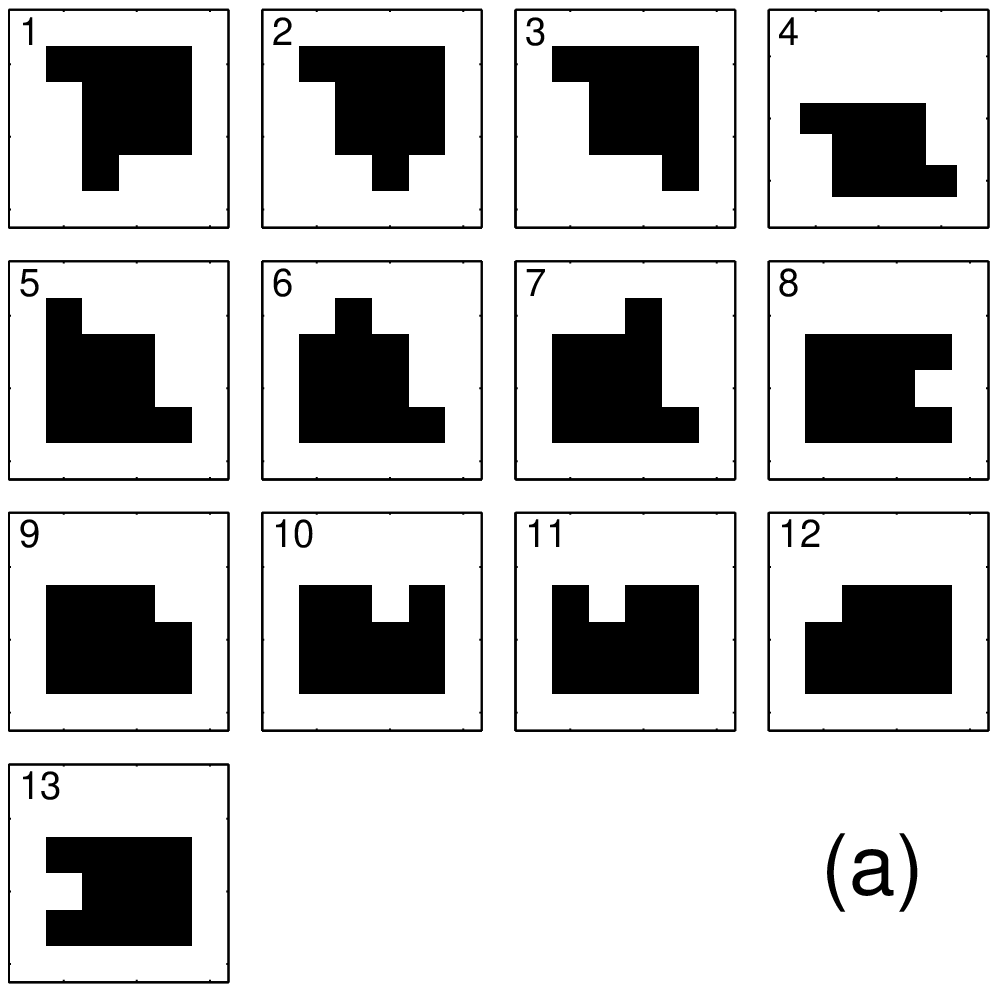}
\includegraphics[width=8.0cm]{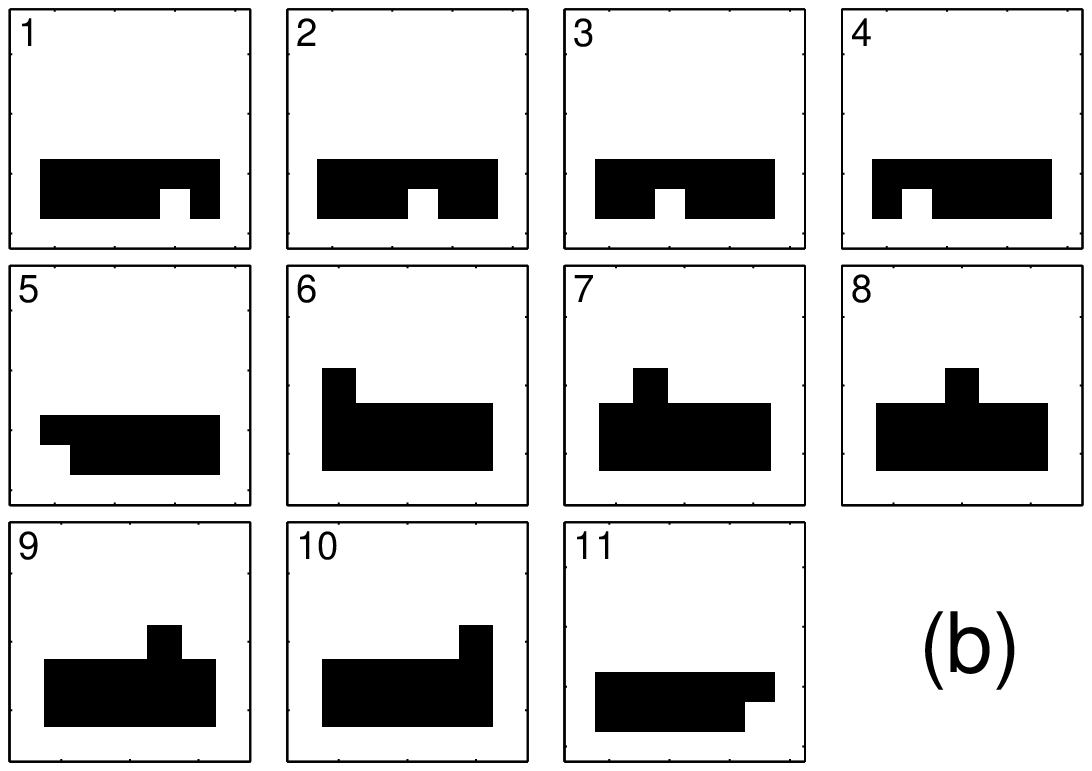}
\caption{The dominating transport cycle for $N=11$ in (a) small and (b) large fields in axis direction.}
\label{fig:sykli_odd_field}
\end{figure}

In Fig.~\ref{fig:sykli_odd_field} we have plotted typical dominating transport cycles for $N=11$ in the static field case.  Two types of cycles were found, one for very small fields (a) and one for large fields (b).  In the small-field cycle, atoms move around the nearest square-shaped island (corresponding to mean width 3). For a large field the shape of the island is flatter (corresponding to mean width 2).  This type of cycle is found for all small odd-$N$ islands in large fields and proceed by breaking only single nearest-neighbor bonds.  This is a similar mechanism as previously proposed being the easiest diffusion pathway for $N=5$ \cite{Voter}.

\begin{figure}
\includegraphics[width=8.0cm]{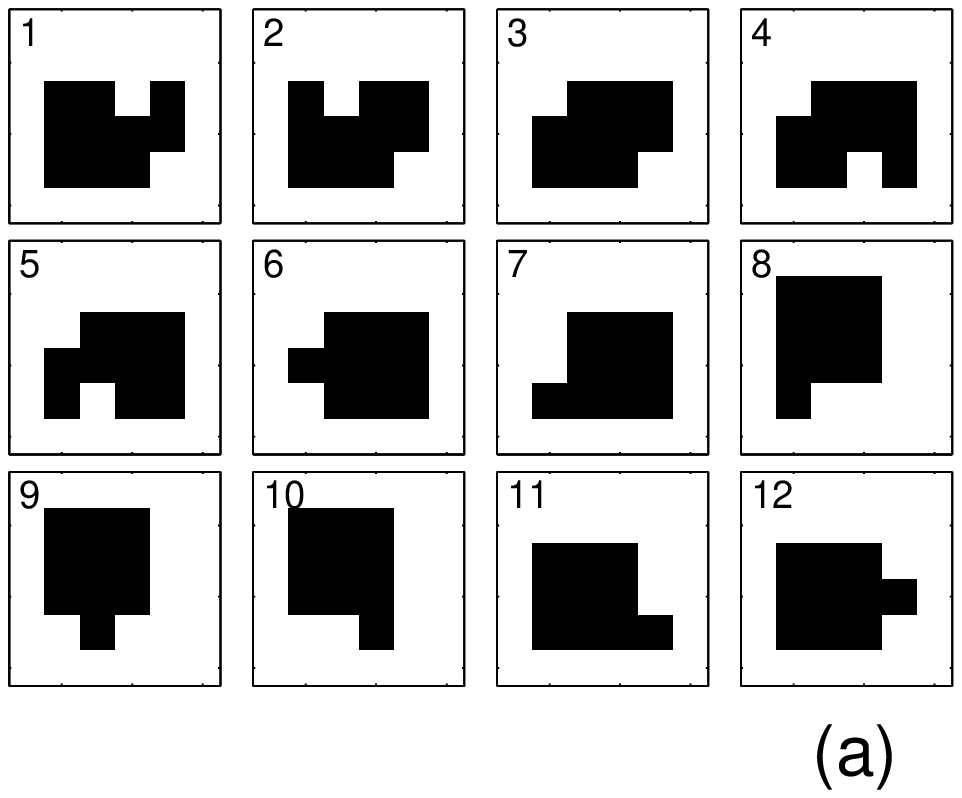}
\includegraphics[width=8.0cm]{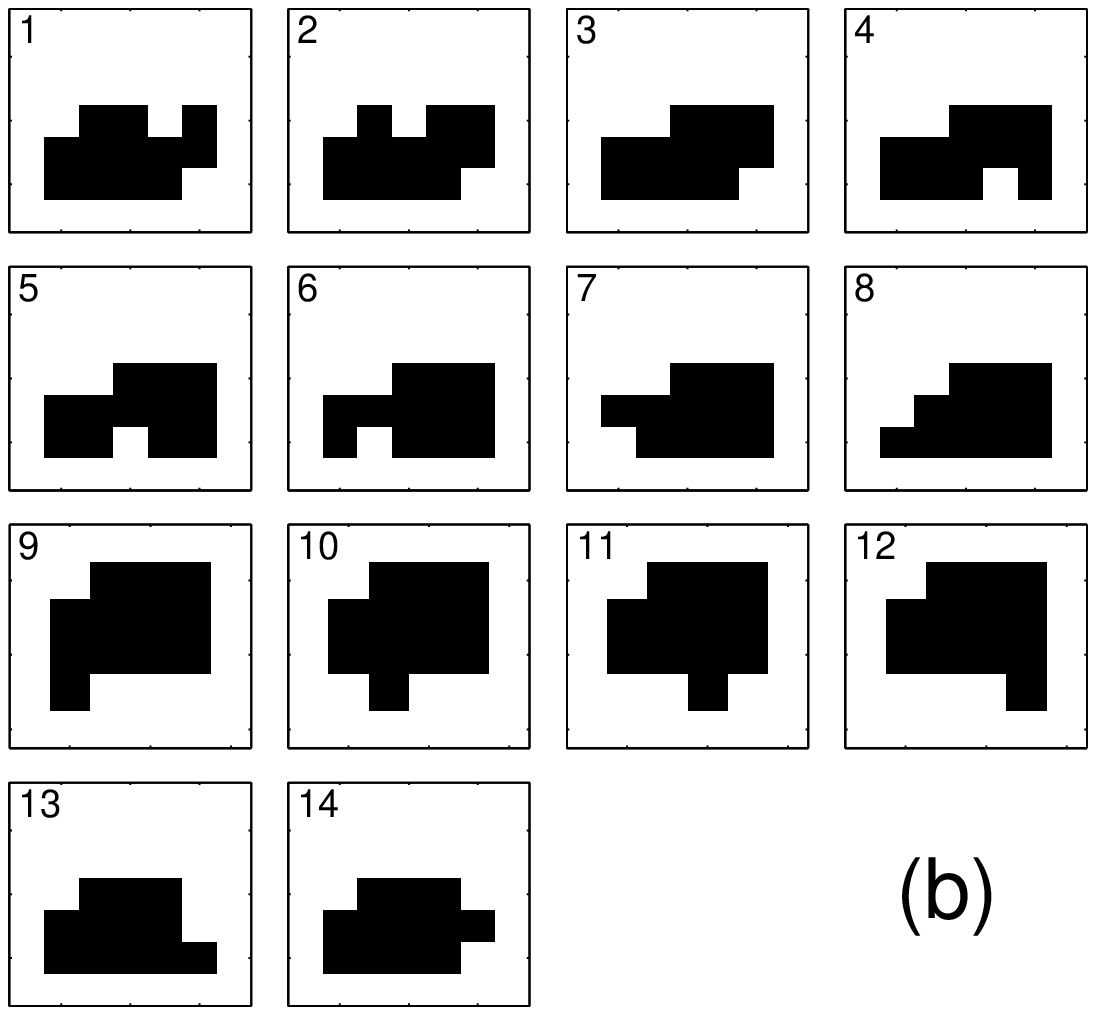}
\caption{The dominating transport cycle for (a) $N=10$ and (b) $N=12$ found for both small and large fields in axis direction.}
\label{fig:sykli_even_field}
\end{figure}

In Fig.~\ref{fig:sykli_even_field} we show the cycles for $N=10$ and $N=12$.  For these even-$N$ islands, only one dominating cycle was found for all fields (note that we only consider fields $E<0.15~{\rm eV}$ because of the trap configurations).  Because of the even number of atoms, similar cycles that were found for $N=11$ in Fig.~\ref{fig:sykli_odd_field}(b) would require breaking of two nearest-neighbor bonds, whereas these two cycles can work with only single bond-breaking transitions.

\begin{figure}
\includegraphics[width=8.0cm]{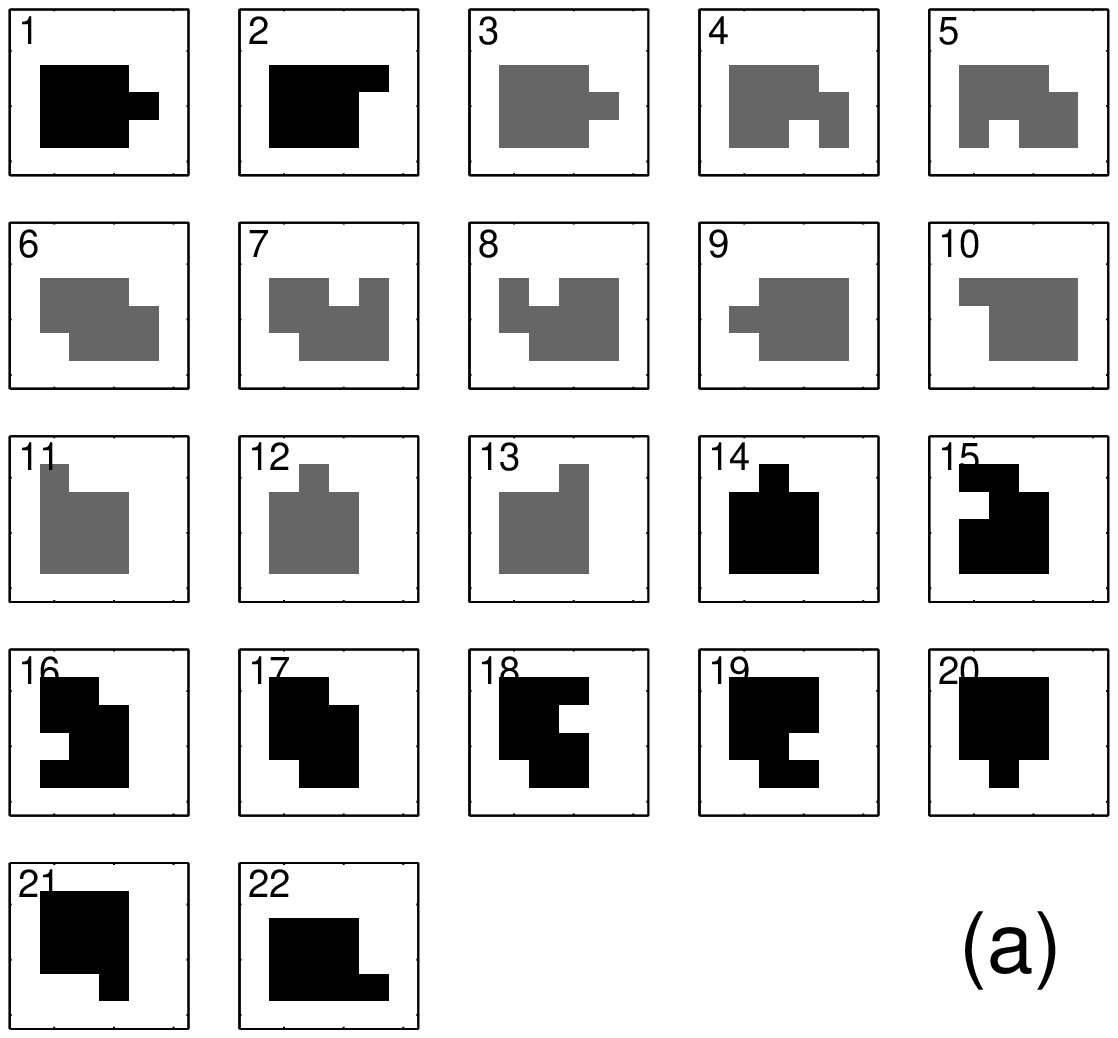}
\includegraphics[width=8.0cm]{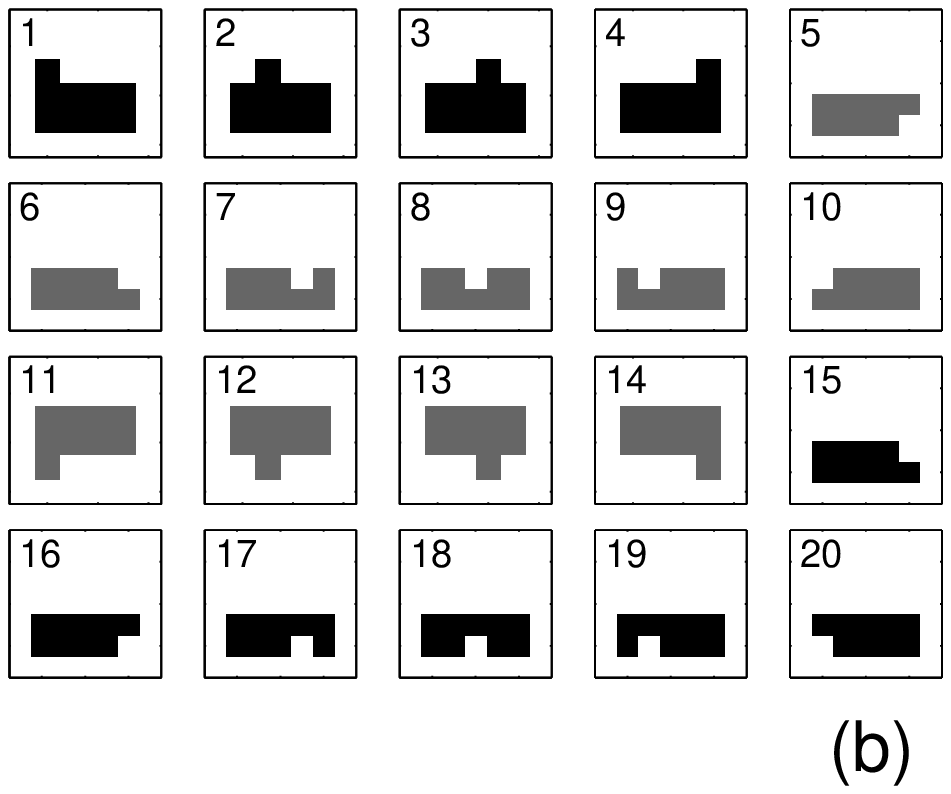}
\caption{A dominating cycles in pulsed field with small $\tau$. (a) $N=10$ with $\gamma=45^\circ$ and $\alpha = 45 \pm 25^\circ$, gray color indicates $\alpha = 20^\circ$ state.  (b) $N=9$ with $\gamma=0^\circ$ and $\alpha = \pm 20^\circ$, gray color indicates $\alpha = -20^\circ$ state.}
\label{fig:sykli_pulsed}
\end{figure}

In Fig.~\ref{fig:sykli_pulsed} we show the dominating cycles in the pulsed field case for $N=9$ and $N=10$ with small $\tau$.  In figure (a) the cycle is shown for $N=9$ using $\gamma=0^\circ$ and $\alpha = \pm 20^\circ$, and in figure (b) for $N=10$, $\gamma=45^\circ$ and $\alpha = 45 \pm 20^\circ$.  With these parameters, the velocity is increased when compared to the static field case (see Sec.~IV.A).  The configurations for $\alpha = -20^\circ$ and $\alpha = 20^\circ$ are shown in gray and the change of potential occurs between the gray and black frames.  The cycles are basically the same as for the static field case except that the barriers for the transitions are lowered due to the pulsed field.  This stochastic-resonance-type mechanism, where the time-scales of two processes are matched, explains the results seen in Sec.~IV.A for the increase of the velocity for small $\tau$.

\section{Discussion}

We have studied the dynamics of single-layer metal-on-metal islands under strong static and time-dependent forces with a continuous-time Monte Carlo (MC) and master equation (ME) methods.  The aim of this work was to study complex non-linear transport phenomena arising when islands are driven out of equilibrium, far beyond the linear response regime.  A semi-empirical model was used and numerical results were presented for Cu atoms on the Cu(001) surface which served as a model system.  Several non-linear effects were identified, most importantly the increase of the velocity by a rotated field and velocity inversions in the electrophoretic ratchet.  By computing the effective Arrhenius parameters and leading relaxation times for small islands using the ME method, non-monotonous behavior was found.  Although the behavior was found to be highly complex, depending strongly on many parameters, such as temperature, field (angle, amplitude and period) and island size, generic behavior could be identified.

First we studied static-field transport up to field strengths corresponding single bond-breaking energy barrier (i.e.~term $E_B$ in Eq.~\eqref{eq:taajuudet}, $0.260~{\rm eV}$ for Cu(001)).  For large fields, many differences arise when compared to equilibrium or linear-response conditions because typical island configurations are no longer nearly square, but are heavily stretched by the field.  A strong even-odd dependence on the island size was found, which has an influence to even large islands (up to $\sim 100$ atoms).  Field amplitudes, wfor which the non-linear behavior first emerges ($E = 0.01 \dots 0.1~{\rm eV}$ for Cu), were found to be especially important for the phenomena reported in this work. In this regime the model is also expected to remain somewhat realistic, based on the observations of the island geometry changes (no rod configurations) and simulations (small island break-up rate).  This is also the regime, where the results for the MC and ME models coincide well.  The direction of the field with respect to the axis was found to have a large effect on the drift.  Introducing a small deviation between the measurement and field angles usually leads to an increased drift.  Although this is expected in the case of the diagonal measurement direction, it was also found in the direction of axis, which purely results from many-particle interactions coupling the internal degrees of freedom with the center-of-mass motion.  The two-maxima structure for the velocity was found as a function of the field angle.

Using the ME method, it was found that the effective barrier computed from the Arrhenius curves depends strongly on the temperature and field strength.  As a result, the dynamics of islands is no longer well described by a single effective energy rate-limiting barrier, which is the case in equilibrium and linear response regime.  The effective barriers are also very different when computed using the effective diffusion coefficient or the drift.  By computing the second-highest eigenvalue of the stochastic generator (the highest one being zero), it was found that the leading relaxation time displays a non-monotonous behavior as a function of the field strength for small islands.  The physical meaning of this is unclear and further studies are needed.

When the periodic time-dependent variation was added to the field, a complex dependence between the velocity and the field period was found.  First we studied the pulsed-field case using symmetrically rotated fields around the measuring direction along the axis and the diagonal.  It was found that the increased velocity was typically produced at the limit of very large period (i.e.~slowly varying field) and maxima/minima were found for smaller periods.  The second type of the field was an electrophoretic ratchet that creates a time-dependent force with a zero mean force.  It was found to produce current inversion phenomena for all large islands.  There are two types of inversion: a genuine time-dependent inversion and an adiabatic inversion for a slowly varying field.  Since current inversions are not possible for a single atom, it is a many-particle effect.  In theory, this type of electrophoretic ratcheting would allow separation of islands based on their size.  In contrast with the velocity increase phenomenon for the pulsed field, current inversion occurs already in very small fields near linear response regime.  One must however note that velocities in the electrophoretic ratchet are very small when compared to velocities for non-zero mean force fields.  

For both types of time-dependent fields, it was found that for certain large field periods (namely for $\tau = 10^4 \dots 10^6$ for temperatures $T = 400 \dots 700\,{\rm K}$), the velocity has a minimum point for large islands.  This time scale corresponds to the process of an atom breaking two nearest neighbor bonds, which is the effective energy barrier process found in this and all previous studies for this model.

Our results indicate that the typical large island behavior begins already for islands with just above $10$ atoms and the small-size effects become much weaker for larger islands.  A similar result was also found in previous studies in equilibrium \cite{Voter}.  For this reason the behavior of the $10$ atom island was found to be somewhat special.  Most phenomena found in this work can already be produced with islands up to $12$ atoms. In general, lowering the temperature tends to make the phenomena such as velocity increase and inversion much stronger at the expense of significantly reducing the absolute velocities.  The current inversion in the electrophoretic ratchet may disappear completely in large temperatures.  This indicates that a large separation in time scales is a required element for these phenomena (at high temperature limit all rates become equal).  Increasing the field amplitude amplifies the velocity increase and inversion up to some point.  Very strong fields however can have a decreasing effect.  Because of this complex dependence on temperature and the field, a data collapse by dimensionless $E/T$ is not possible far from equilibrium, which is in contrast to the linear response regime \cite{Karimi1}.

The ME and MC models were found to be in generally good agreement.  Using suitable parameters, both models were able to reproduce most of the key findings of this paper - especially in small fields.  The largest differences were found for the smallest islands $N<8$, for which the aggressive state reduction (i.e.~the island must be connected via nearest-neighbor bonds) of the ME model appears to have the largest effect.  The vacancy diffusion process was not found to have any significant effect for the ME model. Also the differences between stochastic and deterministic field variation was found to have only a minor effect. One may assume that this is because the time-scale separation of different processes are large, hence the time-scales remain well separated also for the random field periods. 

 By applying the ME model, we were able to investigate large portions of parameter-space with high accuracy, compute effective exponents of the Arrhenius curves and relaxation times, and also identify typical reaction pathways of the islands during transport.  The numerically exact ME method shows its power in making the non-linear effects and their systematics discernible. However, the MC model arguably remains physically more realistic than the ME for the treatment of atoms diffusing around a corner and for the field switching scheme.

Since the barrier structure of our semi-empirical model for the processes on fcc(100) surface is quite generic \cite{Merikoski1}, one can expect similar non-linear phenomena to be present also for other metal-on-metal systems.  As long as distinctive barriers exist, the non-linear transport properties reported here are not limited to any precise values of barriers.  Although our model is simple, it displays a rich variety of phenomena.  This emphasizes the complexity of nonequilibrium many-particle systems and that there is still much to be done in exploring transport in the presence of time-dependent fields.  The simple model does not allow a direct comparison with experimental data, but our conclusions are generic in nature.  By introducing more accurate energetics and adding new microscopic transition types, it is possible that some phenomena disappear while new ones might appear, which we demonstrated by comparing the MC and ME models.  It would be also interesting to study similar properties on other lattice geometries such as close-packed surfaces, with the effect of steps, strain, detachment/attachment processes and other types of driving or interfering forces included.

\acknowledgments

This work was supported by the V\"ais\"al\"a Fund via Finnish Academy of Science and Letters (J.K.).

\end{document}